\documentclass[lettersize,journal]{IEEEtran}
\usepackage{amsmath,amsfonts}
\usepackage{color,multirow}
\usepackage{cite}
\usepackage{algorithm}
\usepackage{algcompatible}
\algblockdefx{FORP}{ENDFORP}[1]%
  {\textbf{for}#1 \textbf{do in parallel}}%
  {\textbf{end for}}
\algblockdefx{Repeat}{Until}[1]%
  {\textbf{repeat}#1}%
  {\textbf{until} converged}
\algblockdefx{If}{Else}[1]%
  {\textbf{if} #1}%
  {\textbf{else}}
  
\usepackage{array}
\usepackage[caption=false,font=normalsize,labelfont=sf,textfont=sf]{subfig}
\usepackage{textcomp}
\usepackage{stfloats}
\usepackage{url}
\usepackage{verbatim}
\usepackage{graphicx}
\usepackage{cite}
\DeclareMathOperator\CE{CE}
\DeclareMathOperator\Mel{Mel}

\begin{document}

\title{Robust One-Shot Singing Voice Conversion}

\author{
    Naoya Takahashi~\IEEEmembership{Member,~IEEE,}, 
    Mayank Kumar Singh~\IEEEmembership{Member,~IEEE,}, 
    Yuki Mitsufuji,~\IEEEmembership{Senior Member,~IEEE}
        % <-this % stops a space
}



\maketitle

\begin{abstract}
Recent progress in deep generative models has improved the quality of voice conversion in the speech domain. However, high-quality singing voice conversion (SVC) of unseen singers remains challenging due to the wider variety of musical expressions in pitch, loudness, and pronunciation. Moreover, singing voices are often recorded with reverb and accompaniment music, which make SVC even more challenging. In this work, we present a robust one-shot SVC (ROSVC) that performs any-to-any SVC robustly even on such distorted singing voices. To this end, we first propose a one-shot SVC model based on generative adversarial networks that generalizes to unseen singers via partial domain conditioning and learns to accurately recover the target pitch via pitch distribution matching and AdaIN-skip conditioning. We then propose a two-stage training method called \textit{Robustify} that train the one-shot SVC model in the first stage on clean data to ensure high-quality conversion, and introduces enhancement modules to the encoders of the model in the second stage to enhance the feature extraction from distorted singing voices. To further improve the voice quality and pitch reconstruction accuracy, we finally propose a hierarchical diffusion model for singing voice neural vocoders. 
Experimental results show that the proposed method outperforms state-of-the-art one-shot SVC baselines for both seen and unseen singers and significantly improves the robustness against distortions.
% Recent progress in deep generative models has improved the quality of voice conversion in the speech domain. However, high-quality singing voice conversion (SVC) of unseen singers remains challenging due to the wider variety of musical expressions in pitch, loudness, and pronunciation. Moreover, singing voices are often recorded with reverb and accompaniment music, which make SVC even more challenging. In this work, we present a robust one-shot SVC (ROSVC) that performs any-to-any SVC robustly even on such distorted singing voices. To this end, we first propose a one-shot SVC model based on generative adversarial networks that learns to accurately recover the target pitch via pitch distribution matching and AdaIN-skip conditioning. We then propose a two-stage training method called \textit{Robustify} that improves the robustness against the distortion. In the first stage, the one-shot SVC model is trained on clean data to ensure high-quality conversion, and in the second stage, enhancement modules are introduced to the encoders of the model to enhance the feature extraction from distorted singing voices. To furhter improve the voice quality and pitch reconstruction accuracy, we finally propose a hierarchical diffusion model for singing voice neural vocoders. 
% Experimental results show that the proposed method outperforms state-of-the-art one-shot SVC baselines for both seen and unseen singers and greatly improves the robustness against the distortions.
\end{abstract}

\begin{IEEEkeywords}
one-shot singing voice conversion, noise robust, neural vocoder, diffusion models
\end{IEEEkeywords}

\section{Introduction}
\renewcommand{\thefootnote}{\fnsymbol{footnote}}
% \footnote[0]{* indicates equal contribution}
\label{sec:intro}
The aim of singing voice conversion (SVC) is to convert a source singing voice into another singer's voice while maintaining the melody and lyrical content of the given source. %applications
SVC has attracted increasing attention due to its potential applications over a wide area including content creation, education, and entertainment.
% Despite progress in speech voice conversion, SVC remains challenging due to a wider variety of pitch range and musical expressions. and is more sensitive to pitch error because singing voices with pitch errors are perceived as off-pitch and fail to maintain the original melody. Many SVC approaches have been proposed to address this problem and have shown promising results \cite{Kobayashi14, Nachmani19, Deng20, Polyak20, Luo20, Zhang20, Liu21, Takahashi21,Liu21FastSVC, Guo22, Zhang22, Zhou22}. 
Despite recent advancements of voice conversion in speech domain, SVC remains challenging owing to following reasons: (1) singing voices have a wider variety of pitch, loudness, and pronunciation owing to different styles of musical expression, which make them more challenging to model; (2) human perception is often sensitive to a singing voice with pitch error because it is perceived as off-pitch and fails to maintain the melodic contents; (3) the scarcity of large-scale clean singing voice datasets hinders generalization of SVC models; and (4) SVC models are prone to distortion of input singing voices.
As a result, many SVC approaches have focused on converting a singing voice into those seen during the training (known as a many-to-many case) in relatively small and clean datasets \cite{Kobayashi14, Nachmani19, Deng20, Polyak20, Luo20, Liu21, Takahashi21,Liu21FastSVC, Guo22, Zhang22, Zhou22,Jayashankar2023}. % TODO: check citations
% Many SVC approaches have been proposed to address this problem and have shown promising results \cite{Kobayashi14, Nachmani19, Deng20, Polyak20, Luo20, Zhang20, Liu21, Takahashi21,Liu21FastSVC, Guo22, Zhang22, Zhou22}. 
% List up related works here.
% However, most of the previous SVC systems focus on converting a singing voice to those seen during the training, known as a many-to-many case (or any-to-many if the source voice can be from unseen singers). 
However, it is often difficult or even impossible to collect a clean singing voice from the target singer in advance. Thus, extending SVC models to unseen target singers (known as a any-to-any case) is an inevitable requirement for many practical applications.
% This hinder the applicability of SVC methods.
Moreover, in many cases a singing voice will be modified with a reverb effect and face interference from music, as a singer will often sing along with accompaniment music. The distortions caused by the interference of music and reverb contaminates the singing voice and hinders the extraction of the acoustic features required for SVC (e.g., pitch, linguistic content, and singer's voice characteristics), thus leading to a severe degradation in the SVC performance.
One way to mitigate this problem is to use music source separation and dereverberation algorithms to remove music and reverb from recordings. However, they often produce non-negligible artefacts, and using such processed samples for input to an SVC system will still considerably degrade the SVC quality. 

In this paper, we propose a robust one-shot singing voice conversion (ROSVC) that robustly generates the singing voice of any target singer from any source singing content even with a distorted voice. The proposed model takes as reference less than ten seconds of the singing voice from a (possibly unseen) target singer and converts the source singer's voice robustly in a one-shot manner even when the singing voice has interference from accompaniment music and modified with the reverb effects.
To this end, we propose three components; (i) a neural network architecture and training framework that enables one-shot SVC with accurate pitch control, (ii) a two-stage training called \textit{Robustify} that improves the robustness of the feature extraction against the distortions of the input singing voice, and (iii) a hierarchical diffusion model for a singing voice vocoder that learns multiple diffusion models at different sampling rates to improve the quality and pitch stability.

Our intensive experiments on the NUS48E, NHSS, and MUSDB18 datasets show that the proposed method outperforms five one-shot VC baselines on both seen and unseen singers and significantly improves the robustness against distortions caused by reverb and accompaniment music. 

% TODO: summary of contributions:
Our contributions are summarized as follows:
\begin{enumerate}
    \item We propose a network architecture and training method for one-shot singing voice conversion to enables the generation of high quality singing voice with accurate pitch recovery.
    \item To consider a more practical and challenging scenario of singing voice containing accompaniment music and reverb, we propose a framework called Robustify that significantly improves the robustness of the SVC model against such distortions.
    \item We further propose a hierarchical diffusion model-based neural vocoder to generate a high-quality singing voice
    \item We conduct extensive experiments using various singing voice datasets and show that the proposed method outperforms state-of-the-art one-shot SVC models and significantly improves the robustness against distortions caused by interference from accompaniment music and reverb.
\end{enumerate}

Part of this work on the hierarchical diffusion model-based neural vocoder was published as a conference paper \cite{Takahashi2022HPG}, that focused on the vocoding task for ground truth acoustic features.  In the paper, we newly propose a novel one-shot SVC that is robust against distortions and investigate the hierarchical diffusion model-based neural vocoder on the challenging SVC scenario. 

Audio samples are available at our website \footnote{\url{https://t-naoya.github.io/rosvc/}\label{link:demo}}.
% $^{\ref{link:demo}}$.

\section{Related works}
\subsection{Singing voice conversion}
Unique challenges in SVC is to accurately recover the target pitch and handle wide variety of pitch, loudness, and musical expression.
Initial SVC approaches tackled the SVC problem by utilizing parallel data \cite{Villavicencio10, Kobayashi14}. Several methods have been proposed to overcome the necessity of the expensive parallel data by using deep generative models such as autoregressive models \cite{Nachmani19, Deng20, Zhang20,Takahashi21}, variational autoencoders \cite{Luo20}, GANs \cite{Polyak20, Guo22,Zhou22,Liu21FastSVC}, and diffusion models \cite{Liu21}. However, they are limited to many(any)-to-many or many-to-one cases and cannot handle unseen target singers. Other approaches leverage a speaker recognition network (SRN) to extract the speaker embeddings from reference audio \cite{Zhang20}. Li et al. \cite{Li22} investigated a hierarchical speaker representation for one-shot SVC. Our approach is different as their training objectives are to reconstruct the input voice from disentangled features and the conversion is done only during inference by changing the speaker embeddings, while in our approach, the input voices are converted during the training so that it does not suffer from training-inference mode gap and more directly constrains the converted samples. Moreover, all previous approaches focus on clean singing voice and are prone to distortions.  In contrast, our work aims at any-to-any SVC on possibly distorted singing voice without parallel data.

\subsection{One-shot voice conversion}
One-shot VC has been actively investigated in the speech domain \cite{chou2019adainvc, Wu20VQVCp, Chen21AGAINVC, Lin21FragmentVC}. AdaIN-VC \cite{chou2019adainvc} uses a speaker encoder to extract speaker embeddings and condition the decoder using adaptive instance normalization (AdaIN) layers. VQVC+ \cite{Wu20VQVCp} extracts speaker-independent content embeddings using vector quantization and utilizes the residual information as speaker information. Fragment-VC \cite{Lin21FragmentVC} utilizes a cross-attention mechanism to use fragments from reference samples to produce a converted voice.
% AdaIN-VC \cite{chou2019adainvc} and AGAIN-VC \cite{Chen21AGAINVC}  use adaptive instance normalization layer to remove the speaker information from the source and condition the model with speaker embeddings. VQVC+ uses vector quantization to extract discrete contents representations and treat the residual information as speaker information. Fragment-VC employs an cross-attention to use fragments from reference samples to produce outputs from phonetic features. 
Although these approaches have shown promising results on speech, they cannot be scaled to singing voices due to their simplicity, as shown in our experiment. 
% Li et al. \cite{Li22} investigated one-shot SVC on clean singing voice. 
\subsection{Noise robust voice conversion}
There have been a few attempts to improve the robustness of speech VC against noise by using a clean--noisy speech pair to learn noise robust representations \cite{Miao20, Mottini21}. These approaches are not directly applicable because our model converts the singer identity during the training and there is no denoised target that we can utilize to train the model via some reconstruction losses. Xie et al. propose leveraging a pre-trained denoising model and directly using noisy speech as a target signal \cite{Xie22}.  In these studies, environmental sounds are used as noise, which is uncorrelated to voice. 
In contrast, we consider reverb and accompaniment music as noise, which can be more challenging because accompaniment music often contains a similar harmonic structure to the singing voice, and strong reverb effects on the singing voice make robust feature extraction more difficult.
\subsection{Neural vocoder}
\label{sec:vocoder}
Voice conversion models often operate in acoustic feature domains to efficiently model the mapping of speech characteristics. Neural vocoders are often used for generating high-quality waveform from acoustic features \cite{Aaron2016WN, Mehri2017SampleRNN, Prenger2019WaveGlow, Ping20WaveFlow, Yamamoto20PWG, Kong21DiffWave}. A number of generative models have been adopted to neural vocoders including autoregressive models \cite{Aaron2016WN,Mehri2017SampleRNN,Kalchbrenner2018WaveRNN}, generative adversarial networks (GANs) \cite{Donahue2019WaveGAN,Kumar2019MelGAN,Yamamoto20PWG,Kong2020HiFiGAN}, and flow-based models \cite{Prenger2019WaveGlow, Ping20WaveFlow}.
Recently, diffusion models \cite{Song2019NCSN, Ho2020DDPM} have been shown to generate high-fidelity samples in a wide range of areas \cite{Lu2022cDPM} and have been adopted for neural vocoders in the speech domain \cite{Kong21DiffWave,Chen2021WaveGrad}. Although they are efficiently trained by maximizing the evidence lower bound (ELBO) and can produce high-quality speech data, the inference speed is relatively slow compared to other non-autoregressive model-based vocoders as they require many iterations to generate the data.
To address this problem, PriorGrad \cite{Lee22PriorGrad} introduces a data-dependent prior, i.e., Gaussian distribution with a diagonal covariance matrix whose entries are frame-wise energies of the mel-spectrogram. As the noise drawn from the data-dependent prior provides an initial waveform closer to the target than the noise from a standard Gaussian, PriorGrad achieves faster convergence
and inference with a superior performance. SpecGrad \cite{Koizumi22SpecGrad} further improves the prior by incorporating the spectral envelope of the mel-spectrogram to introduce noise that is more similar to the target signal. 

However, we found that state-of-the-art neural vocoders provide insufficient quality when they are applied to a singing voice. In this work, we address this problem by proposing a hierarchical diffusion model.

\begin{figure*}[t]
  \centering
  \includegraphics[width=0.78\linewidth]{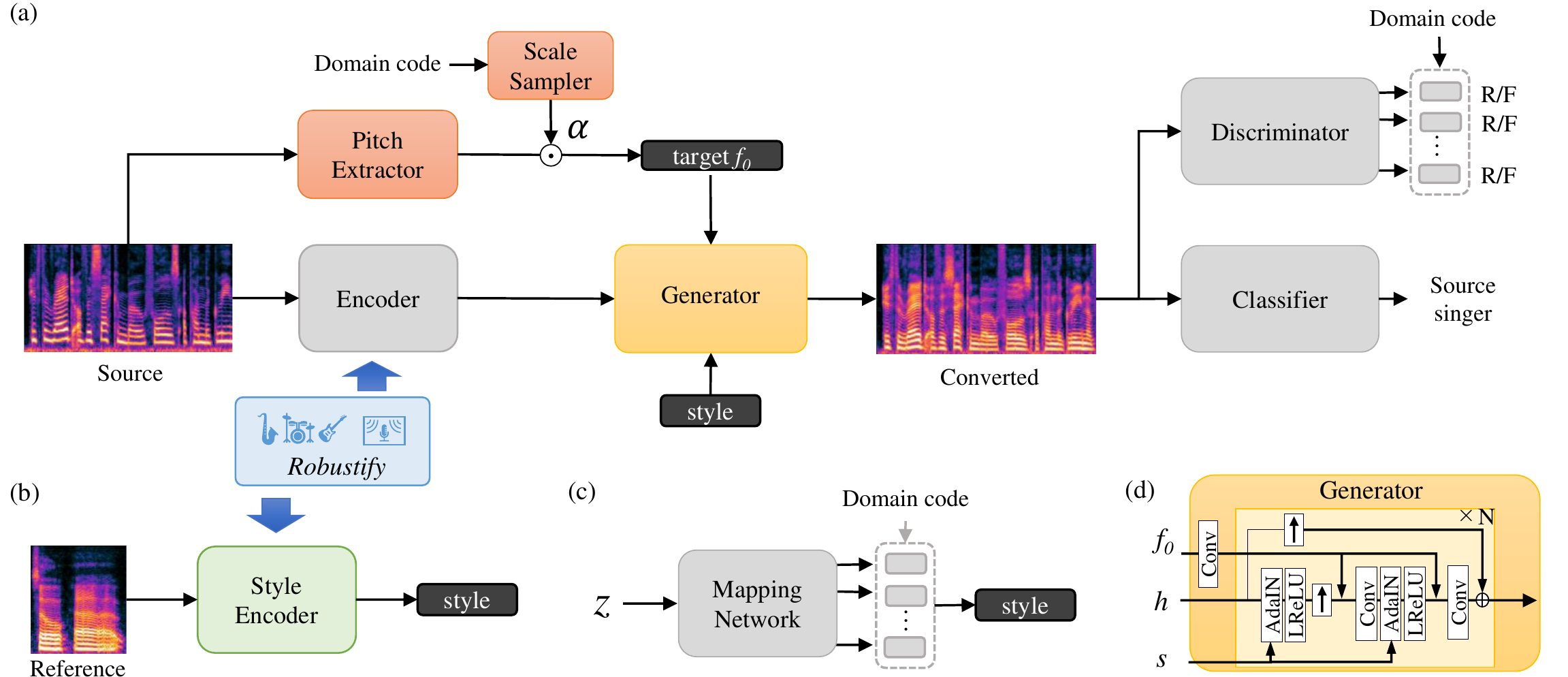}
  \caption{Robust one-shot SVC framework. (a) The generator is conditioned on the style vector and the target $f_0$ to be reconstructed in the converted sample. The target $f_0$ is obtained by scaling the source $f_0$ to match the target distribution based on the domain statistics. (b),(c) Unlike the mapping network, the style encoder is domain-independent and yet trained to fool the domain-specific discriminators. The encoders are refined to improve the robustness against input distortions in the second stage of the training.}
  \label{fig:overview}
\end{figure*}
% TODO: Encoder -> content Encoder
% TODO: Separate figure for architecture

\section{Proposed One-shot SVC Framework}
\label{sec:osvcframework}
We base our SVC model on the generative adversarial network (GAN)-based voice conversion model called StarGANv2-VC \cite{Li21}. Although StarGANv2-VC yields excellent sample quality in speech voice conversion, it has several limitations when it applied to SVC: (i) It is limited to the many(any)-to-many case and cannot generate the singing voice of unseen speakers. (ii) When it is applied to a singing voice, the converted voice often sounds off-pitch. (iii) It does not have pitch controllability, which is important for SVC as converted singing voices are often played with an accompaniment and thus the pitch should be aligned with the accompaniment music. (iv) The conversion quality is severely degraded when the singing voices contain reverb and accompaniment music. To address these problems, we first extend StarGANv2-VC by introducing a domain-independent style encoder for enabling one-shot voice conversion (Sec.\ref{sec:osvc}) and introduce AdaIN-skip pitch conditioning to enable accurate pitch control (Sec.\ref{sec:pitch}). We then introduce a two-stage training framework called \textit{Robustify} to improve the robustness of the feature extraction against the distortions (in Sec. \ref{sec:robustify}). Finally, we introduce a hierarchical diffusion model for the singing voice vocoder to enable high-quality signing voice waveform generation (Sec.s \ref{sec:HDM}).

\subsection{One-shot SVC framework with domain-specific and domain-independent modules}
\label{sec:osvc}
The overview of the proposed one-shot SVC framework is shown in Figure \ref{fig:overview}. The generator $G(h, f^{trg}_0, s)$ converts the source mel-spectrogram $X_{src}$ into a sample in the target domain $X_{trg}$ based on an encoder output $h=E(X_{src})$, target fundamental frequency (F0) $f^{trg}_0$, and style embedding $s$, where the domain in our case is the singer identity. $h$ is a time-varying feature expected to contain linguistic contents while $s$ is a global feature expected to contain the singer's voice characteristics. The discriminator $D$ consists of shared layers followed by domain-specific heads to classify whether the input is real or fake on each target domain via adversarial loss
\begin{equation}
L_{adv}=\mathbb{E}_{X,f^{trg}_0,s}[\log D(X,y_{src}) -\log D(G(h, f^{trg}_0, s),y_{trg})],
\end{equation}
where $y_{trg}\in\mathcal{Y}$ denotes the target domain.
Although the domain-specific discriminator helps make the generator's outputs realistic and similar to the target domain, we further promote the conversion by introducing an additional classifier $C$. The classifier $C()$ takes as input the generated sample and is trained to identify the source domain $y_{src}$ via classification loss $L_{cl}(y)$ while the generator is trained to fool the classifier via the adversarial classification loss $L_{ac}(y)$:
\begin{eqnarray}
    L_{cl} =\mathbb{E}_{X,f^{trg}_0,s}[\CE(C(G(h, f^{trg}_0, s)),y_{src})],\\
    L_{ac} =\mathbb{E}_{X,f^{trg}_0,s}[\CE(C(G(h, f^{trg}_0, s)),y_{trg})],
    \label{eq:ce}    
\end{eqnarray}
where $\CE$ denotes the cross-entropy loss.
The style embedding $s$ is obtained by either the style encoder or the mapping network. Given the target domain $y_{trg}$, the mapping network $M$ transforms a random latent code $z \sim \mathcal{N}(0, 1)$  into the style embedding as $s=M(z, y_{trg})$. 
In the original StarGANv2 \cite{Li21,Choi20StarGANv2}, the mapping network, style encoder, and discriminator have domain-specific projection heads to enable the model to easily handle domain-specific information and focus on diversity within the domain. However, this architecture limits the conversion within the pre-defined domains, which is many-to-many SVC in our case. To enable any-to-any one-shot SVC, we propose using a domain-independent style encoder $S(X)$ while keeping the domain-specific heads for the mapping network and discriminator. By doing so, the style encoder does not require the domain code and can then transform any singer's voice during the inference time while the domain-specific mapping network and discriminator still guide the model to handle the domain-specific information. We empirically demonstrate that this design does not deteriorate the conversion quality compared to the original many-to-many model. Note that the mapping network is not utilized for inference, as our goal is one-shot SVC, but it isstill useful for guiding the model to learn domain specific characteristics and the diversity within the domains.

\subsection{Pitch conditioning}
\label{sec:pitch}
 Unlike speech voice conversion, accurate pitch reconstruction is essential for SVC to maintain the melodic content. Although the StarGANv2-VC model in \cite{Li21} uses the $f_0$ feature extracted from the \textit{source} by an F0 estimation network to guide the generation, the output is only weakly constrained to have a normalized F0 trajectory similar to that of the source. Therefore, the absolute pitch of the converted sample is uncontrollable, and we found that it often produces off-pitch singing voices. To address this problem, we propose conditioning the generator on the \textit{target} pitch and force the converted sample to reconstruct the exact conditioned pitch:
\begin{equation}
    L_{f0} = \mathbb{E}_{X, f^{trg}_0\sim\mathcal{F}^y, s}[\| f_0^{trg} - F(G(h, f^{trg}_0, s))\|_1],
    \label{eq:f0}
\end{equation}
where $F(X)$ is the F0 network that estimates the F0 value of the input $X$ in Hertz. 
Ideally, the target F0 should be sampled from the F0 distribution of the target domain $f_0^{trg}\sim\mathcal{F}^y$ to fool the domain-specific discriminator and the classifier while maintaining the melodic content. To achieve this, we stochastically scale the source F0 by $\alpha=\mathcal{P}(y_{src},y_{trg})$ during the training, where $\mathcal{P}$ is the scale sampler for matching the F0 distribution of the scaled source F0 with that of the F0 in the target domain. We pre-compute the F0 histogram for the singers in the training set and sample the scale value so that the mean of the scaled pitch $mean(\alpha f_0^{src})$ matches the target singer's F0 distribution during the training. During the inference time, we simply compute the scale value as a ratio of the mean F0 values $\alpha=mean(f_0^{trg})/mean(f_0^{src}) $.

The generator architecture for the pitch conditioning is shown in Figure \ref{fig:overview} (d). In StarGANv2-VC \cite{Li21}, pitch features are concatenated with the encoder output $h$ and fed to the AdaIN layers, but we found that the model struggles to recover the absolute target pitch with this architecture because the absolute value of the target pitch tends to be lost due to the instance normalization in the AdaIN operation. Therefore, we propose skipping the AdaIN layers for pitch conditioning by concatenating the F0 features obtained from the $f_0$ by a convolution layer to the features after AdaIN and nonlinearities. This enables us to preserve the absolute pitch information and accurately recover the pitch.
% Unlike the style embedding, in which AdaIN layers are used for conditioning, we skip the AdaIN layers for pitch conditioning, and F0 features obtained from the $f_0$ by a convolution layer are concatenated to the features after AdaIN and nonlinearities. We found that skipping the AdaIN layer is essential for reconstructing the target pitch because the absolute value of the target pitch tends to be lost due to the instance normalization in the AdaIN operation. The model struggles to recover the absolute target pitch when we concatenate the F0 feature at the beginning, as done in \cite{Li21}.

\subsection{Training objectives}
\label{sec:obj}
The goal of our one-shot SVC training is to learn mapping functions $E,G,S$ that convert a sample $X\in\mathcal{X}_{\mathcal{Y}_{src}}$ from the source singer $\mathcal{Y}_{src}$ into a sample $\hat{X}=G(E(X),f_0^{trg},S(X^{trg}))\in\mathcal{X}_{\mathcal{Y}_{trg}}$ in the target singer's domain $\mathcal{Y}_{trg}$ specified by $X^{trg}\in\mathcal{X}_{\mathcal{Y}_{trg}}$ while maintaining the lyric content in $X$ and recovering the target pitch $f_0^{trg}$ in $\hat{X}$. 
To achieve this, we additionally incorporate the following losses.
To maintain the linguistic contents, the speech consistency loss,
\begin{equation}
L_{asr}=\mathbb{E}_{X,s}[\|A(X) - A(G(h(X), f_0, s))\|_1],
\end{equation}
is employed by using an automatic speech recognition model $A$. 
To ensure that the singer style is extracted from the reference singer and refracted to the generated singing voice, the style reconstruction loss,
\begin{equation}
L_{sty}=\mathbb{E}_{X,s}[\|s - S(G(h(X), f_0, s))\|_1],
\end{equation}
is introduced.
We also enforce the model to generate diverse samples with different style embeddings $s_1,s_2$ by maximizing the style diversification loss,
\begin{equation}
L_{ds}=\mathbb{E}_{X, s_1, s_2}[\|G(h(X), f_0, s_1) - G(h(X), f_0, s_2)\|_1].
\end{equation}
Lastly, the cycle consistency loss is employed to preserve all other features of the input as
\begin{equation}
L_{cyc}=\mathbb{E}[\|X - G(h(G(h(X), f'_0, s')),f_0,s)\|_1].
\end{equation}
The overall objectives for modules $E,G,S,M,D,C$ are given by
\begin{equation}
    \begin{split}
    L_{E,G,S,M} &= L_{adv} + \lambda_{ac}L_{ac} + \lambda_{f_0}L_{f_0} + \lambda_{sty}L_{sty} \\
    & \quad - \lambda_{ds}L_{ds} + \lambda_{asr}L_{asr} + \lambda_{cyc}L_{cyc}, 
    \\
    L_{D,C} &= -L_{adv} + \lambda_{cl}L_{cl},
    \end{split}
    \label{eq:objective}
\end{equation}
where $\lambda_{ac},\lambda_{f_0},\lambda_{sty},\lambda_{ds},\lambda_{asr}\lambda_{cyc},\lambda_{adv},\lambda_{cl}$ are the weight parameters.
Unlike StarGANv2-VC, the proposed method does not need to know the target domain $\mathcal{Y}_{trg}$ during the inference but instead extracts the singer embedding from a short reference sample. With sufficient diversity in the training data, the singer embedding space can cover many different styles, and the style encoder learns to extract the style from a reference sample. This enable us to apply our method to a sample from an unseen speaker.

\begin{figure}[t]
  \centering
  \includegraphics[width=\linewidth]{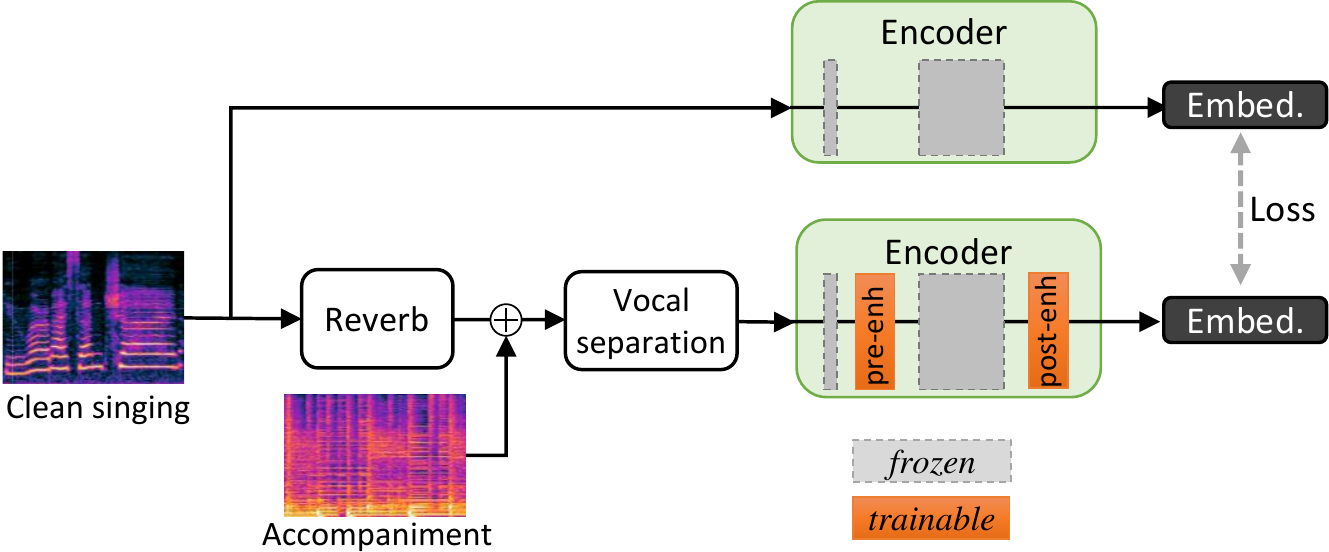}
  \caption{Overview of Robustify training scheme. The same scheme is applied for the style encoder.
  After the encoder is trained on clean data, we freeze the parameters and insert pre- and post-enhancement blocks in the encoder. The distortions are simulated by convolving RIR, adding accompaniment, and separating the singing voice by a source separation model. We train only enhancement blocks by minimizing the distance between embeddings extracted from clean and distorted samples.}
  \label{fig:robustify}
\end{figure}

\section{Robustify: Two-stage training for \\improving robustness}
\label{sec:robustify}
Although the model described above achieves high-quality one-shot SVC on clean data, the quality of a generated sample will be degraded if the source and/or target samples have distortions such as the interference of accompaniment music or reverb effect. To overcome this problem, we introduce a two-stage training method called \textit{Robustify} that aims to improve the quality of SVC against the distorted inputs. Since it is fairly feasible to collect clean singing voices from some singers during the development, in the first stage, we train the one-shot SVC model on the clean dataset (as described in Sec. \ref{sec:osvcframework}). 
Once the model is trained, we freeze the model parameters and insert two new enhancement blocks, namely, pre- and post-enhancement blocks, in the encoder and the style encoder as shown in Figure \ref{fig:robustify}. Let $\tilde{E}_\theta$ and $\tilde{S}_\psi$ denote the encoder and style encoder in which the enhancement blocks are added, where $\theta$ and $\psi$ are the parameters of the enhancement blocks in $\tilde{E}$ and $\tilde{S}$, respectively. 
% The distorted sample $X'$ is simulated by distortion function $\zeta: x\rightarrow X'$.
The distortion is simulated by convolving a room impulse response (RIR) $r$ to the singing voice and adding instrumental music $m$. We then leverage a pre-trained vocal separation network $\xi$ called D3Net \cite{Takahashi21D3Net} to extract the vocals from the mixture, which alleviates the significant characteristic change of the input caused by the interference of music.
The distorted sample $x'$ is obtained by
\begin{equation}
    X'=\zeta(x)=\Mel(\xi(x*r+m)),
\end{equation}
where $x$ denotes a time domain singing voice, $*$ denotes convolution, $\Mel$ the mel feature extraction function, and $X=\Mel(x)$.
We train the newly added enhancement blocks by minimizing the L1 distance between the embeddings extracted from clean samples by the original frozen encoder and the outputs obtained by the encoders with enhancement blocks from the distorted samples,
\begin{equation}
    \begin{split}
    L_{ro}^{\theta} = \mathbb{E}_{x,r,m}[\|E(X)- \tilde{E}_{\theta}(X')\|_1 + \lambda\|E(X)- \tilde{E}_{\theta}(X)\|_1],
    \\
    L_{ro}^{\psi} = \mathbb{E}_{x,r,m}[\|S(X)- \tilde{S}_{\psi}(X')\|_1 + \lambda\|S(X)- \tilde{S}_{\psi}(X)\|_1],
    \end{split}
    % \label{eq:robustify}
    \notag
\end{equation}
% where $\tilde{E}$ and $\tilde{S}$ denote the encoder and style encoder in which the enhancement blocks are added, $\theta$ and $\psi$ are the parameters of the enhancement blocks in $\tilde{E}$ and $\tilde{S}$, respectively, $X^{dist}$ is the mel-spectrogram of the distorted sample, and $\lambda$ is the weight.
where the second and fourth terms are regularization terms to ensure that the enhancement blocks do not deteriorate the performance of clean samples, and $\lambda$ is the weight.
The pre-enhancement block is placed after the initial convolution layer while the post-enhancement block is introduced at the end of the encoder. We find that this architecture provides a lower $L_{ro}$ loss than the pre- or post-enhancement only settings. Each enhancement block consists of two residual blocks that have the same architecture as those in the encoder.
% The distortions are simulated by convolving a room impulse response (RIR) $r$ to the singing voice and adding instrumental music $m$. We then leverage a pre-trained vocal separation network $\xi$ called D3Net \cite{Takahashi21D3Net} to extract the vocals from the mixture, which alleviates the significant characteristic change of the input caused by the interference of music.
% The distorted sample as obtained by
% \begin{equation}
%     X'=\zeta(x)=\Mel(\xi(x*r+m)),
% \end{equation}
% where $x$ denotes a time domain singing voice, $$*$ denotes convolution, $\Mel$ the mel feature extraction function, and $X=\Mel(x)$.
Note that the output of the separation network is not perfect and still contains reverb, interference, and artefacts introduced by the separation network, so using the separated vocals directly without the enhancement blocks still degrades the SVC quality considerably, as shown in our experiment (Sec. \ref{sec:ex_dist}). 
% Evidence of design choice?
% We refer the proposed second stage feature enhancement scheme as \textit{Robustify}.

The proposed two-stage training has several advantages. First, since we freeze the encoders and generator trained on the clean data, we can guarantee that the training on the distorted data in the second stage will not deteriorate the quality of the SVC on clean data by skipping enhancement blocks. Thus, the model can harness the advantage of clean data when the test samples are also known to be clean.  Moreover, the Robustify scheme greatly reduces the maximum computation and memory cost required for training compared to when the entire model
is trained to handle distorted samples from scratch. This is because we can omit the modules required for synthesizing the distorted samples (i.e., reverb and vocal separation modules) from the first stage while omitting the generator, discriminator, classifier, and mapping network from the second stage. This makes it possible to train a robust model on a single GPU and apply data augmentation on the fly.

\begin{figure*}[t]
  \centering
  \includegraphics[width=0.75\linewidth]{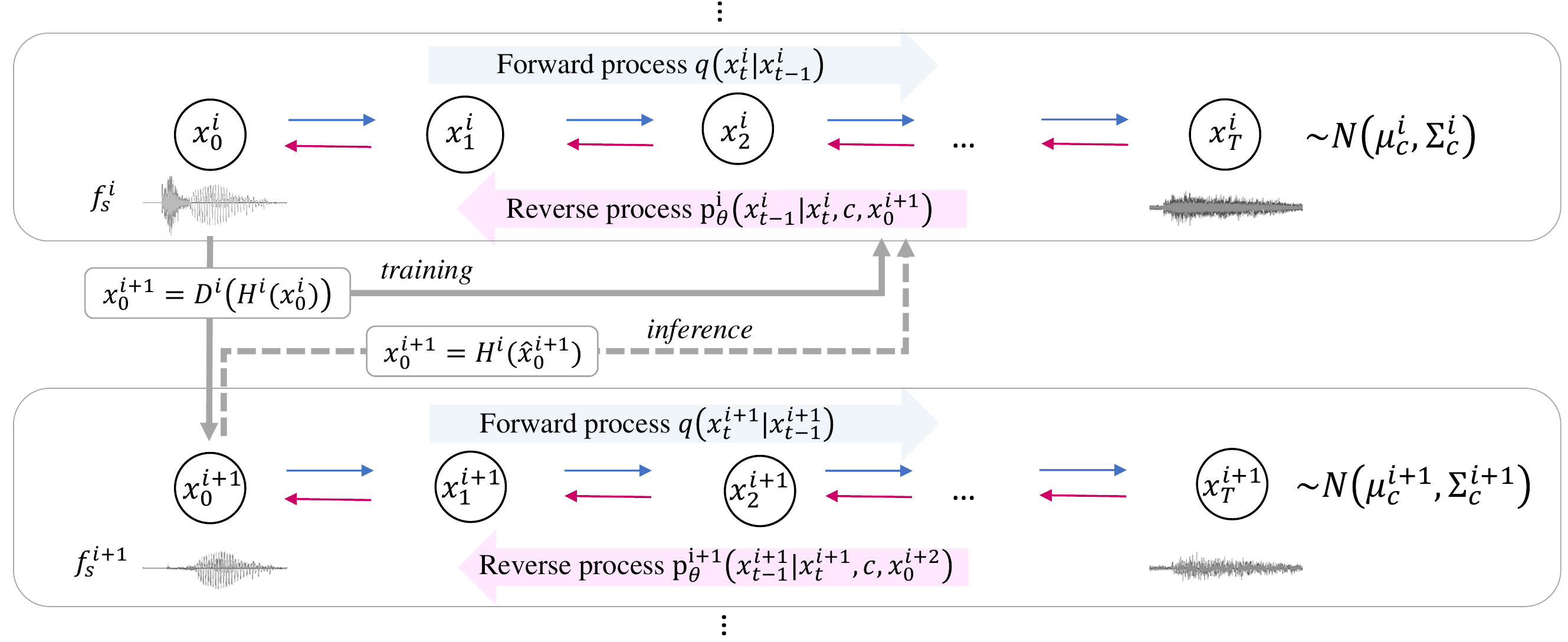}
  \caption{Overview of proposed hierarchical diffusion model combined with PriorGrad \cite{Lee22PriorGrad}. Diffusion models are trained at multiple sampling rates $f_s^1>\cdots>f_s^N$ independently. Each diffusion model is conditioned on acoustic features $c$ and data at the lower sampling rate $x_0^{i+1}$. During inference, the anti-aliasing filter $H^i$ is applied to the data generated at the lower sampling rate and used for conditioning.}
  \label{fig:overviewHDM}
\end{figure*}

\section{Hierarchical diffusion models for singing voice vocoder}
\label{sec:HDM}
% Recently, neural vocoders are commonly used to recover high quality waveform from the mel-spectrogram. As in \cite{Li21}, we first tried Parallel WaveGAN (PWG) \cite{Yamamoto20PWG} vocoder and found that it tends to produce unnatural shakes in pitch and volume when the singing voice uses expressive technique such as vibrato. We retrain the PWG on singing data and observe a little improvement. However, we found larger improvement when we use a diffusion-based model called PriorGrad \cite{Lee22PriorGrad}, thus, we chose it for our experiments.
As described in Sec. \ref{sec:vocoder}, neural vocoders are utilized to recover the waveform from the mel-spectrogram. We found that state-of-the-art neural vocoders \cite{Aaron2016WN, Yamamoto20PWG, Kong2020HiFiGAN} still struggle to generat high-quality singing voices and tend to produce unnatural shakes in pitch when the source singing voice contains vibrato, even after we trained models on singing voice data.
To overcome this problem, we propose a hierarchical diffusion model for a singing voice vocoder. The proposed hierarchical diffusion model learns multiple diffusion models at different sampling rates in parallel. The diffusion model at the lowest sampling rate can focus of the recovery of the accurate fundamental frequency components while the models at higher sampling rates can leverage the outputs of the models at the lower sampling rate and focus on adding high frequency components. In the following subsections, we first summarize diffusion-based neural vocoders and then describe the proposed hierarchical diffusion model.

\subsection{Denoising diffusion probabilistic models (DDPM)}
\label{sec:ddpm}
DDPMs consist of two Markov chains called the \textit{forward} and \textit{reverse} processes. The \textit{forward process} gradually diffuses the data $x_0$ into a standard Gaussian $x_T$, as
% $x_0 \in \mathbb{R}^D$
\begin{equation}
    q(x_{1:T}|x_0) =\prod^T_{t=1} q(x_t|x_{t-1}),
    \label{eq:fw}
\end{equation}
where $q(x_t|x_{t-1}):= N(x_t;\sqrt{1-\beta_t}x_{t-1}, \beta_t \textbf{I})$ is a transition probability at a time-step $t$ that adds a small Gaussian noise on the basis of a noise schedule $\beta_t\in\{\beta_1,\cdots,\beta_T\}$. This formulation enables us to directly sample $x_t$ from $x_0$ as: 
% $q(x_t|x_0)=\mathcal{N}(\sqrt{\bar{\alpha}_t}x_0, (1-\bar{\alpha}_t)\textbf{I})$
\begin{equation}
x_t=\sqrt{\bar{\alpha}_t}x_0 + \sqrt{(1-\bar{\alpha}_t)}\epsilon,
    \label{eq:xt}
\end{equation}
where $\alpha_t=1-\beta_t$, $\bar{\alpha}_t=\prod_{s=1}^t\alpha_s$, and $\epsilon\sim\mathcal{N}(\textbf{0},\textbf{I})$. The noise schedule is designed to make $\bar{\alpha}_T$ very small so that $x_T$ converges to $\epsilon\sim\mathcal{N}(\textbf{0},\textbf{I})$.
The \textit{reverse process} gradually transforms the prior noise $p(x_T)=\mathcal{N}(x_T;\textbf{0},\textbf{I})$ to data as:
\begin{equation}
    p(x_{0:T}) =p(x_T)\prod^T_{t=1} p_\theta(x_{t-1}|x_t),
    \label{eq:rev}
\end{equation}
% where $p_\theta(x_{t-1}|x_t)$ is a transition propability that corresponds to the reverse of $q(x_t|x_{t-1})$ and is modeled by a deep neural network parameterized by $\theta$.
where $p_\theta(x_{t-1}|x_t):=\mathcal{N}(x_{t-1};\mu_\theta(x_t,t),\sigma_\theta^2(x_t,t)\textbf{I})$ is a transition probability that corresponds to the reverse of $q(x_t|x_{t-1})$ and is modeled by a deep neural network parameterized by $\theta$.
As shown in\cite{Ho2020DDPM}, $p_\theta(x_{t-1}|x_t)$ can be given by $\mathcal{N}(\mu_\theta(x_t,t), \sigma_\theta(x_t,t))$ with
\begin{eqnarray}
    \mu_\theta(x_t,t)=\frac{1}{\sqrt{\alpha_t}}(x_t-\frac{\beta_t}{\sqrt{1-\bar{\alpha}_t}}\epsilon_\theta(x_t,t)),
\end{eqnarray}
and $\sigma_\theta^2(x_t,t)=\frac{1-\bar{\alpha}_{t-1}}{1-\bar{\alpha}_t}\beta_t$, where $\epsilon_\theta(x_t,t)$ is a deep neural network that predics the noise $\epsilon$ added at time $t$ in \eqref{eq:xt}. The model $\epsilon_\theta(x_t,t)$ can be learned by maximizing the ELBO:
\begin{equation}
    ELBO =-\sum_{t=1}^T\kappa_t\mathbb{E}_{x_0,\epsilon}[||\epsilon-\epsilon_\theta(x_t,t)||^2]+C,
    \label{eq:elbo}
\end{equation}
where  $\kappa_t=\frac{\beta_t}{2\alpha(1-\bar{\alpha}_{t-1})}$ for $t>1$, $\frac{1}{2\alpha}$ for $t=1$ and C is a constant. 
% As suggested in \cite{Ho2020DDPM}, many followup works instead use a simplified loss function by setting $\kappa_t=1$ \cite{Ho2020DDPM,Kong21DiffWave,Lee22PriorGrad}. DDPM have been adopted to neural vocoders by conditioning the noise estimation network on acoustic features $c$ as $\epsilon_\theta(x_t,c,t)$ \cite{Chen2021WaveGrad,Kong21DiffWave}. Starting from the noise sampled from the prior $x_T$, the DDPM-based vocoders iteratively denoise the signal $x_t$ on the basis of the condition $c$ to obtain the corresponding waveform $x_0$.

Many followup works use a simplified loss function instead by setting $\kappa_t=1$ \cite{Ho2020DDPM,Kong21DiffWave,Lee22PriorGrad} as suggested in \cite{Ho2020DDPM}. DDPMs have also been adopted in neural vocoders \cite{Chen2021WaveGrad,Kong21DiffWave}.
% by modeling waveform and conditioning the noise estimation network on acoustic features $c$ as $\epsilon_\theta(x_t,c,t)$ \cite{Chen2021WaveGrad,Kong21DiffWave}. 
The DDPM-based vocoders iteratively denoise the signal $x_t$ on the basis of the acoustic feature $c$ using the noise estimation network  $\epsilon_\theta(x_t,c,t)$ to obtain the corresponding waveform $x_0$ starting from the noise sampled from the prior $x_T$.
\subsection{PriorGrad}
% Although the standard Gaussian prior in DDPMs provides a simple solution without any assumption on the target data, it requires many steps to obtain high-quality data, which hinder efficient training and sampling. To improve the efficiency in the neural vocoder case, PriorGrad \cite{Lee22PriorGrad} uses an adaptive prior $\mathcal{N}(\textbf{0},\mathbf{\Sigma_c})$, where the diagonal variance $\mathbf{\Sigma_c}$ is computed from a mel-spectrogram $c$ as $\mathbf{\Sigma_c}= diag[(\sigma_0^2, \cdots, \sigma_L^2)]$ and $\sigma_i^2$ is a normalized frame-level energy of the mel-spectrogram at the $i$th sample. The loss function is modified accordingly to 
% \begin{equation}
% L=\mathbb{E}_{x_0,\epsilon,t}[||\epsilon-\epsilon_\theta(x_t,c,t)||^2_{\Sigma^{-1}}],
% \end{equation}
% where $||\mathbf{x}||^2_{\Sigma^{-1}}=\mathbf{x}^\top\Sigma^{-1}\mathbf{x}$. Intuitively, as the power envelope of the adaptive prior is closer to that of the target signal than that of the standard Gaussian prior, the diffusion model can require fewer time steps to converge and be more efficient.   

The standard Gaussian prior in DDPMs offers a simple solution without assuming anything about the target data. However, it requires many steps to obtain high-quality data, thus hindering efficient training and sampling. PriorGrad \cite{Lee22PriorGrad} reduces the number of steps by using an adaptive prior $\mathcal{N}(\textbf{0},\mathbf{\Sigma_c})$  in the case of neural vocoders, where $\sigma_i^2$ in the diagonal variance $\mathbf{\Sigma_c}$ is a frame-level normalized energy of the mel-spectrogram at the $i$th sample. $\mathbf{\Sigma_c}$ is calculated from a mel-spectrogram $c$ as $\mathbf{\Sigma_c}= diag[(\sigma_0^2, \cdots, \sigma_L^2)]$. The loss function is accordingly modified to
\begin{equation}
L=\mathbb{E}_{x_0,\epsilon,t}[||\epsilon-\epsilon_\theta(x_t,c,t)||^2_{\Sigma^{-1}}],
\end{equation}
where $||\mathbf{x}||^2_{\Sigma^{-1}}=\mathbf{x}^\top\Sigma^{-1}\mathbf{x}$. In comparison to the standard Gaussian prior, the power envelope of the adaptive prior is closer to that of the target signal. This results in the diffusion model requiring fewer reverse diffusion steps to converge.

\subsection{Hierarchical diffusion probabilistic model}
\label{sec:HDPM}
% Although PriorGrad shows promising results on speech data, we found that the quality is unsatisfactory when it is applied to a singing voice, possibly due to the wider variety in pitch, loudness, and musical expressions such as vibrato and falsetto. To tackle this problem, we propose to improve the diffusion model-based neural vocoders by modeling the singing voice in multiple resolutions. An overview is illustrated in Figure \ref{fig:overview}. Given multiple sampling rates $f_s^1>f_s^2>\cdots>f_s^N$, the proposed method learns diffusion models at each sampling rate independently. 

% The reverse processes at each sampling rate $f_s^{i}$ are conditioned on common acoustic features $c$ and the data at the lower sampling rate $f_s^{i+1}$ as $p^i_\theta(x^i_{t-1}|x^i_t, c, x^{i+1}_0)$ except the model at the lowest sampling rate, which is conditioned only on $c$.
% During the training, we use the ground truth data $x_0^{i+1}=D^i(H^i(x^i_0))$ to condition the noise estimation models $\epsilon^i_\theta(x^i_t, c, x^{i+1}_0,t)$, where $H^i(.)$ denotes the anti-aliasing filter and $D^i(.)$ denotes the downsampling function for the signal at the sampling rate of $f_s^i$. 

While PriorGrad has demonstrated promising results with speech data, we found that the quality of generated voices is unsatisfactory when adopted for singing voices. This discrepancy may arise from the increased diversity in pitch, loudness, and musical expressions such as vibrato and falsetto. To address this issue, we propose improving diffusion model-based neural vocoders by modeling the singing voice in multiple resolutions. An overview of the proposed hierarchical diffusion model is depicted in Figure \ref{fig:overviewHDM}. Given the sampling rate $f_s$, the proposed method learns diffusion models at $N$ different sampling rates $f_s=f_s^1>f_s^2>\cdots>f_s^N$ independently.
The model at the sampling rate $f_s^{i}$ is conditioned on common acoustic features $c$, and the data at the lower sampling rate $f_s^{i+1}$ as $p^i_\theta(x^i_{t-1}|x^i_t, c, x^{i+1}_0)$ except for the model at $f_s^N$, which is conditioned only on $c$.
To avoid dependency on other models, the noise estimation models are conditioned using the ground truth data $x_0^{i+1}=D^i(H^i(x^i_0))$ as $\epsilon^i_\theta(x^i_t, c, x^{i+1}_0,t)$ during the training, where $D^i(.)$ denotes the down-sampling function and $H^i(.)$ denotes the anti-aliasing filter for the signal at the sampling rate of $f_s^i$. 

\begin{algorithm}
\caption{Training of Hierarchical PriorGrad}\label{alg:train}
\begin{algorithmic}
\State \textbf{Given:} $f_s^1>\cdots>f_s^N$
\FORP { $i=1,\cdots,N$}
\Repeat
\State $x^i_0 \sim q^i_{data}, \epsilon^i\sim\mathcal{N}(\textbf{0},\mathbf{\Sigma_c}),t\sim\mathcal{U}([0,\cdots,T])$
\State $x_0^{i+1}=D^i(H^i(x^i_0))$ if $i<N$; else $x_0^{i+1}=$~\textit{Null}
% \If $i=N$
% \State $x_0^{i+1}=$~\textit{Null}
% \Else
% \State ~~~~$x_0^{i+1}=D^i(H^i(x^i_0))$    ~~\Comment{Downsample data}    
\State $x^i_t=\sqrt{\bar{\alpha}_t}x^i_0 + \sqrt{(1-\bar{\alpha}_t)}\epsilon$
\State $L=||\epsilon^i - \epsilon_\theta(x^i_t, c, x^{i+1}_0,t)||^2$
\State Update the model parameter $\theta$ with $\nabla_\theta L$
\Until
\ENDFORP
\end{algorithmic}
\end{algorithm}

Since the noise $\epsilon$ is linearly added to the original data $x_0$ as in \eqref{eq:xt} and the model has direct access to the ground truth lower-sampling rate data $x^{i+1}_0$, the noise prediction for low-frequency components from $x^i_t$ and $x^{i+1}_0$ can be simplified by avoiding the complicated acoustic feature-to-waveform transformation. This enables the model to dedicate its capacity more on the transformation of high-frequency components. 
% As the model has access to the oracle lower-sampling rate data $x^{i+1}_0$, it can more easily predict low-frequency components of the linearly added noise $\epsilon$ to the original data $x_0$ as in \eqref{eq:xt} from $x^i_t$ and $x^{i+1}_0$ by avoiding the complicated acoustic feature-to-waveform transformation. By conditioning the model on $x^{i+1}_0$, it can dedicate its parameters more on the transformation of the high-frequency components.
As the data at lower sampling rates ($f_s^N=6$ kHz in our experiments) become much simpler to model compared to the original sampling frequency $f_s^1$, the models at lower sampling rates can more easily model the low-frequency components, which is crucial for accurate pitch recovery of a singing voice.

% During inference, we start by generating the data at the lowest sampling rate $\hat{x}_0^N$ and progressively generate the data at the higher sampling rate $\hat{x}_0^i$ by using the generated sample $\hat{x}_0^{i+1}$ as the condition. 
% In practice, we found that directly using $\hat{x}_0^{i+1}$ as the condition often produces noise around the Nyquist frequencies of each sampling rate, $\frac{f_s^2}{2},\cdots,\frac{f_s^N}{2}$, as shown in Figure \ref{fig:filter}~(a). 
% This is due to the gap between the training and inference mode;  the ground truth data used for training $x_0^{i+1}=D^i(H^i(x^i_0))$ do not contain a signal around the Nyquist frequency owing to the anti-aliasing filter and the model can learn to directly use the signal upto the Nyquist frequency, while the generated sample used for inference $\hat{x}_0^{i+1}$ may contain some signal around there due to the imperfect predictions and contaminate the prediction at a higher sampling rate.

\begin{figure}[t]
  \centering
  \includegraphics[width=0.7\linewidth]{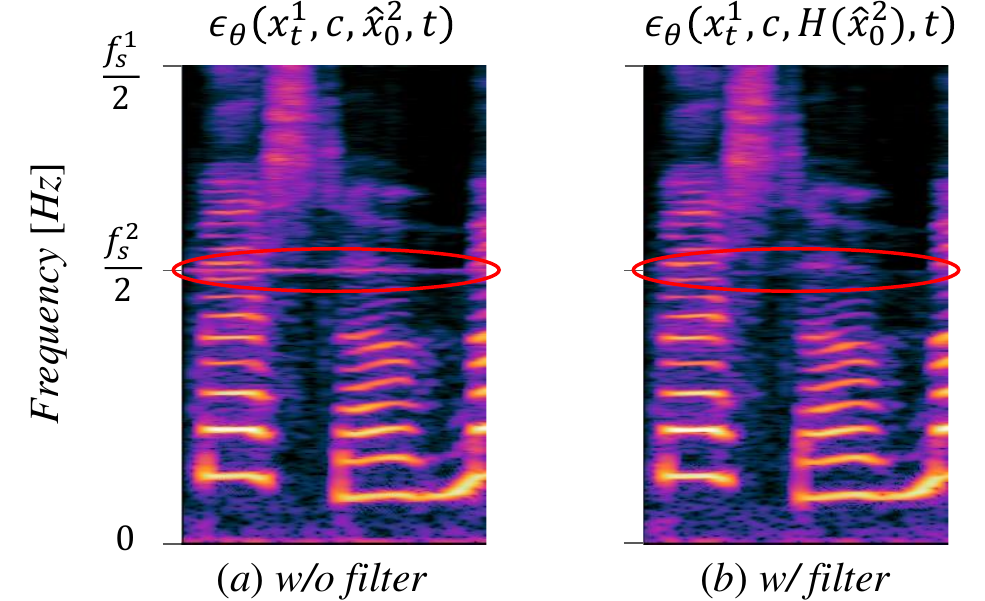}
  \caption{Effect of anti-aliasing filter in the case of $N=2, f_s^1=24000, f_s^2=6000$. Noise is produced around the Nyquist frequency when we directly use the generated waveform at the lower sampling rate without anti-aliasing filter.}
  \label{fig:filter}
\end{figure}

For inference, we first generate the data at the lowest sampling rate $f_s^N$, $\hat{x}_0^N$, and progressively generate the data for the higher sampling rates $\hat{x}_0^i, (i=N-1,\cdots,1)$ by conditioning the model $\epsilon^i_\theta$ on the generated sample $\hat{x}_0^{i+1}$. 
As shown in our experiment \ref{sec:ex_hpg_condition}, the proposed model indeed uses the $\hat{x}_0^i$ to generate the lower-frequency components of $\hat{x}_0^{i+1}$ while the higher-frequency components are generated on the basis of $c$. This mode shift occurs around the Nyquist frequency of the conditioned signal at the lower-sampling rate.
In practice, there is a discrepancy between the conditioning signal used in the training and inference modes: namely, the ground truth data used for training $x_0^{i+1}=D^i(H^i(x^i_0))$ do not contain a signal around the Nyquist frequency owing to the anti-aliasing filter and the model while the generated sample used for inference $\hat{x}_0^{i+1}$ may contain some signal around there due to the imperfect predictions. This discrepancy can produce noise around the Nyquist frequency at the inference time because the model can rely on the fact that the conditioned signal does not contain any signal around the Nyquist frequency during the training, and small noise around the Nyquist frequency in  $\hat{x}_0^{i+1}$ can be unexpectedly magnified, as shown in Figure \ref{fig:filter}~(a). 
% directly using $\hat{x}_0^{i+1}$ as the condition often produces noise around the Nyquist frequencies of each sampling rate, $\frac{f_s^2}{2},\cdots,\frac{f_s^N}{2}$, as shown in Figure \ref{fig:filter}~(a). 
% In practice, the discrepancy introduced during inference by conditioning the model on $\hat{x}_0^{i+1}$ instead of $\x_0^{i+1}$ often produces noise around the Nyquist frequency of $f_s^i$, $\frac{f_s^2}{2}$, for all $i$, as shown in Figure \ref{fig:filter}~(a). 
% The discrepancy between the inference and training mode arises because of the lack of any signal in $x_0^{i+1}=D^i(H^i(x^i_0))$ around the Nyquist frequency in the training data owing to the anti-aliasing filter and the model can learn to directly use the signal upto the Nyquist frequency, while during inference, due to the imperfect prediction from the previous model, the generated sample $\hat{x}_0^{i+1}$ can contain some signal around the Nyquist frequency which contaminates the generation of $\hat{x}_0^i$.
To address this problem, we propose applying the anti-aliasing filter to $\hat{x}_0^{i+1}$ to condition the noise prediction model to generate $\hat{x}_0^i$ as 
\begin{equation}
\hat{\epsilon} = \epsilon^i_\theta(x^i_t, c, H(\hat{x}^{i+1}_0),t).
\label{eq:infer}
\end{equation}
Applying the anti-aliasing filter matches the frequency characteristics and removes the noise around the Nyquist frequencies, as shown in Figure \ref{fig:filter}~(b). The training and inference procedures for the combination with PriorGrad are summarized in Algorithms \ref{alg:train} and \ref{alg:infer}, respectively.

Note that the proposed hierarchical diffusion model can be combined with many types of diffusion models, such as DiffWave \cite{Kong21DiffWave}, PriorGrad \cite{Lee22PriorGrad}, and SpecGrad \cite{Koizumi22SpecGrad}.

\begin{algorithm}
\caption{Inference of Hierarchical PriorGrad}\label{alg:infer}
\begin{algorithmic}
\State \textbf{Given:} $f_s^1>\cdots>f_s^N$
\State $\hat{x}_0^{N+1}=$~\textit{Null}
\FOR { $i=N,N-1,\cdots,1$}
\State $x^i_T \sim\mathcal{N}(\textbf{0},\mathbf{\Sigma_c})$
\FOR { $t=T,T-1,\cdots,1$}
\State $\mathbf{z}\sim\mathcal{N}(\textbf{0},\mathbf{\Sigma_c})$ if $T>1$; else $\mathbf{z}=0$
\State $x^i_{t-1} = \frac{1}{\sqrt{\alpha_t}}(x^i_t-\frac{\beta_t}{\sqrt{1-\bar{\alpha}_t}}\epsilon_\theta(x^i_t,c,H(\hat{x}^{i+1}_0),t))+\sigma_t\mathbf{z}$
\ENDFOR
\ENDFOR ~~$x_0=x_0^1$
\end{algorithmic}
\end{algorithm}

\subsection{Network architecture}
% As in PriorGrad \cite{Lee22PriorGrad}, we also base our model architecture on DiffWave \cite{Kong21DiffWave}. The network consists of $L$ residual layers with bidirectional dilated convolution and repeated dilation factors. The layers are grouped into $m$ blocks, and each block consists of $l=\frac{L}{m}$ layers with dilation factors of $[1,2,\cdots,2^{l-1}]$ (Please refer to \cite{Kong21DiffWave} for more details). 
% DiffWave and PriorGrad use $L=30, l=10$ to cover large receptive fields. Our approach can take advantage of modeling in the lower sampling rates and a smaller network can cover the long signal length in seconds. Thus, we reduce the size to $L=24, l=8$ for all models in different sampling rates to mitigate the increase of computational cost due to the multi-resolution modeling. As shown in our experiments, this hyperparameter provides nearly the same computational cost as the original diffusion model when $N=2$. 

We use the base model architecture on DiffWave \cite{Kong21DiffWave}, as in PriorGrad \cite{Lee22PriorGrad}. The network architecture consists of $L$ residual layers of bidirectional dilated convolution having repeated dilation factors. The layers are grouped into $m$ blocks with each block having a dilation factor of $[1,2,\cdots,2^{l-1}]$ where $l=\frac{L}{m}$ (please refer to \cite{Kong21DiffWave} for more details).
 Our approach can take advantage of modeling in the lower sampling rates and a shallower network can cover the long signal length in seconds. In contrast to DiffWave and PriorGrad, which use $L=30, l=10$ to cover 256ms of receptive fields at 24 kHz, the proposed model can cover 320ms of receptive fields with $L=24, l=8$ for all models at different sampling rates and $f_s^N=6$kHz. As shown in our experiments, this mitigates the increase of computational cost due to the multi-resolution modeling and provides nearly the same computational cost as the original diffusion model when $N=2$.

% Using the same network architecture for all sampling rates effectively changes the receptive field of the models depending on the sampling rate, as illustrated in Figure \ref{fig:rf}. At the lower sampling rate, the model covers a longer time period and focuses on low-frequency components while it covers a shorter time period and focuses on high-frequency components at the higher sampling rate. This design matches our intention of the hierarchical diffusion model because the models are expected to directly use the conditioned data at the lower sampling rate $x_0^{i+1}$ upto the Nyquist frequency $\frac{f_s^{i+1}}{2}$ and focus on transforming the acoustic features to the waveform at the high-frequency.

\begin{figure}[t]
  \centering
  \includegraphics[width=0.8\linewidth]{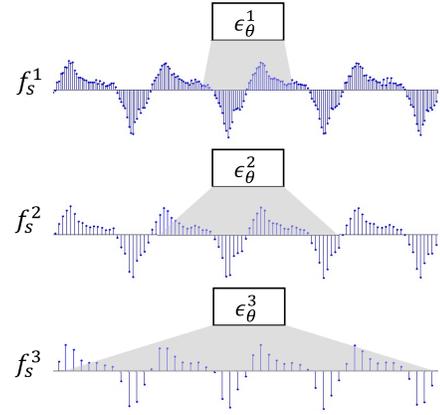}
  \caption{Receptive field at different sampling rates. The same architecture covers a longer time period at lower sampling rates.}
  \label{fig:rf}
\end{figure}

As illustrated in Figure \ref{fig:rf}, the receptive field of the models effectively changes by using the same network architecture for all sampling rates. For the higher sampling rate, the model covers a short time period and focuses on high-frequency components, while at a lower sampling rate, it covers a longer time period and focuses on low-frequency components. Our intention with the hierarchical diffusion model matches this design because the models are expected to use the conditioned lower sampling rate data $x_0^{i+1}$ upto the Nyquist frequency $\frac{f_s^{i+1}}{2}$ and focus on transforming the acoustic features to the high-frequency components of the waveform. 

\begin{table*}[t]
% \begin{minipage}{10.7cm}
    \caption{\label{tab:obj} {\it Objective evaluation of one-shot SVC methods on Clean \textit{dataset}. 
    * indicates minimum PMAE.
    % We report the minimum PMAE for the models that do not have pitch controlability (indicated with *).
    }}
    \vspace{2mm}
    \centering{
      \footnotesize
    \begin{tabular}{c | c c c | c c c} 
    \hline
    \multirow{2}{*}{Model} & \multicolumn{3}{c|}{Seen} & \multicolumn{3}{c}{Unseen}\\
        & PFD $\downarrow$ & Identity $\uparrow$ & PMAE $\downarrow$ & PFD $\downarrow$ & Identity $\uparrow$ & PMAE $\downarrow$\\
    \hline\hline
    Ground Truth & -  & 0.87  & -  & - & 0.87  & -\\
    \hline
    VQVC+ \cite{Wu20VQVCp}& 0.99  & 0.76  & 65.0$^*$ & 1.04 & 0.76 & 49.6$^*$\\
    FragmentVC \cite{Lin21FragmentVC}& 0.70  & 0.77 & 61.8$^*$ & 0.55  & 0.75 & 48.6$^*$\\
    AdaINVC \cite{chou2019adainvc} & 0.76 & 0.80 & 64.7$^*$  & 0.67 & 0.82 & 59.4$^*$\\
    AdaINVC+f0 & 0.73 & 0.82 & 44.8 & 0.60 & 0.79 & 26.3\\
    UCDSVC\cite{Polyak20}+enc & 0.62 & 0.85  & 12.6  & 0.45 &  0.85 & 12.0 \\
    \hline
    ROSVC (Ours) & \textbf{0.54}  & \textbf{0.89}  & \textbf{8.6}  & \textbf{0.37} & \textbf{0.87} & \textbf{9.0}\\
    \hline
    \end{tabular}
    }
% \end{minipage}
% \hfill
\end{table*}

\section{Experimental procedures}
\subsection{Dataset}
% \textbf{Dataset}\hspace{2mm}
We use the NUS48E\cite{Duan13} and NHSS\cite{Sharma21NHSS} corpora to train and test our ROSVC model. NUS48E consists of 12 singers (six male and six female), four songs for each singer, while NHSS consists of 10 singers (five male and five female), ten songs each. We use ten singers from NUS48E and eight singers from NHSS for development and the remaining four for the evaluation on unseen singers. 
The development set is randomly split into 90\%--5\%--5\% for training--validation--test sets. 
For the accompaniment music data used in \textit{Robustify}, we use MUSDB18 \cite{sisec2018}, which includes 100 and 50 professionally recorded songs, for the {\it Dev} and {\it Test} sets, respectively. 
We use {\it bass, drums, other} tracks to form accompaniments.
% and \textit{vocals} track is used. 
Twenty RIRs are generated using a reverb plug-in to simulate the reverb effect. We split them into 80\%-20\% for the train-test sets. 
For the training of the HPG vocoder, we also use an internal corpus consists of eight singers (five male and three female), from 50 to 80 songs each. 
All data are resampled to 24kHz.
\subsection{Training}
% \\
% \textbf{Training}\hspace{2mm}
% We base our ROSVC model on StarGANv2 \cite{Li21,Choi20StarGANv2} and replace the domain-dependent heads of the style encoder to a single head, add pitch conditioning as described in Sec.\ref{sec:pitch}, and add enhancement blocks described in Sec.\ref{sec:robustify}.
We use an 80-band mel spectrogram at log-scale as the acoustic feature. We set the FFT size to 2048 and hop size to 300.
In the first stage, the ROSVC model is trained for 150 epochs with a batch size of 16, and a learning rate of 0.0001 with the AdamW optimizer. We set $\lambda_{cl}=0.1$, $\lambda_{f0}=\lambda_{cyc}=5$, $\lambda_{sty}=\lambda_{ds}=1$, and $\lambda_{asr}=10$. Exponential moving averages over parameters \cite{Yazici2019EMA} of all modules except D are employed as in \cite{Choi20StarGANv2}.
During the second stage of the training, namely \textit{Robustify}, we randomly select the RIR and convolve it with the clean singing voice to form a reverberant signal. To augment the data, the reverberant singing voice is mixed with the original clean signal as $\gamma x*r + (1-\gamma)x$, where $x$ denote the clean signal, $r$ the RIR, and $\gamma \in [0, 1]$ the random scalar value. Accompaniment music is generated by randomly scaling instrument tracks ({\it bass, drums, other}) with a range of [0.75, 1.25] and mixing them. We freeze the model parameters trained in the first stage and train the enhancement blocks only for 150 epochs with a batch size of 16, and a learning rate of 0.0001 with the AdamW optimizer as described in Sec. \ref{sec:robustify}.  

For the HPG vocoder training, we follow the settings used in \cite{Kong21DiffWave, Lee22PriorGrad}: the models are trained for 1M steps with a batch size of 16, learning rate of $2\times10^{-4}$, and Adam optimizer \cite{Kingma2015Adam}. 
We apply the proposed hierarchical diffusion model to PriorGrad \cite{Lee22PriorGrad} and evaluate two models: the 2-stage Hierarchical PriorGrad (HPG-2) with $(f_s^1,f_s^2) = (24k, 6k)$ and the 3-stage Hierarchical PriorGrad (HPG-3) with $(f_s^1,f_s^2,f_s^2) = (24k,12k,6k)$. The inference noise schedule is set to [0.0001, 0.001, 0.01, 0.05, 0.2, 0.5] with $T_{infer}=6$ as in \cite{Lee22PriorGrad}. 
% We followed the publicly available implementation\footnote{\url{https://github.com/microsoft/NeuralSpeech}\label{fn:pg}}.

\section{Results: One-shot SVC}
\subsection{Comparison with state-of-the-art models on clean data}
\label{sec:ex_osvc}
\textbf{Objective evaluation}\hspace{2mm}
% \\
% \textbf{Evaluation metrics}\hspace{2mm}
We randomly select singers and songs for sources and targets and generate 400 8s samples  for each seen-to-seen and unseen-to-unseen singer case for each model.
We use three metrics for objective test: phonetic feature distance (PED), identity, and pitch mean absolute error (PMAE). PED is used to evaluate the linguistic content preservation; we extract features from the source and converted samples using the encoder of the end-to-end speech recognition system \cite{Kim17HybridASR} provided in ESPNet \cite{espnet} and calculate the l2 distance. Identity is the cosine similarity of d-vectors extracted from a speaker verification model
\cite{Resemblyzer}
% \footnote{\url{https://github.com/resemble-ai/Resemblyzer}} 
as in \cite{Wang21}. We use the PMAE between the target F0 and the F0 extracted from the converted sample. For models that do not have explicit target pitch conditioning, we scale the source F0 to produce the minimum PMAE and report the minimum value (mPMAE).  

% We consider five baselines for one-shot SVC. Three one-shot voice conversion methods, namely Fragment VC \cite{Lin21FragmentVC}, VQVC+ \cite{Wu20VQVCp}, and AdaINVC \cite{chou2019adainvc} are adopted to the SVC. 
% % We use publicly available code and  train the models on the same singing voice data as the proposed model. 
% % Since AdaINVC uses the adaptive instance norm for conditioning as our model, 
% We also evaluate a AdaINVC model extended with the proposed target pitch conditioning method. Finally, we also consider a SVC approach in \cite{Polyak20} as the baseline. Since \cite{Polyak20} is for many-to-many senario, we extend it to one-shot SVC by replacing the singer look up table with the same style encoder as the proposed model. 
% \subsection{Objective evaluation}
We first compare the proposed method with the five baselines using objective metrics on the clean dataset to evaluate the one-shot SVC capability. Three one-shot voice conversion methods, namely, FragmentVC \cite{Lin21FragmentVC}, VQVC+ \cite{Wu20VQVCp}, and AdaINVC \cite{chou2019adainvc} are adopted for the SVC.
We also evaluate an AdaINVC model extended with the proposed target pitch conditioning method. Finally, we also include an SVC approach, UCDSVC \cite{Polyak20} as a baseline. Since UCDSVC is originally for many-to-many scenarios, we extend it to one-shot SVC by replacing the singer look-up table with the same style encoder as that of the proposed model. 

The results for seen-to-seen and unseen-to-unseen singer cases are shown in Table \ref{tab:obj}. 
% The simple adaptation of one-shot VC approaches \cite{Lin21FragmentVC, Wu20VQVCp, chou2019adainvc} 
VQVC+, FragmentVC, and AdaINVC have high PMAE values, indicating that directly using one-shot VC approaches for SVC fails to preserve melodic contents as they do not involve an explicit pitch conditioning mechanism.  Adding the pitch conditioning to AdaINVC improves the PMAE, but it remains relatively high. The extension of the UCDSVC \cite{Polyak20} performs best among the baselines. 
The proposed method outperforms all baselines in all metrics, indicating that the converted samples maintain phonetic and melodic contents well and that their singer identities are closer to the target singers. 
% \begin{table}[t]
%     \caption{\label{tab:ablation} {\it Ablation study on seen singers. $^*$ denotes the mPMAE.}}
%     \vspace{2mm}
%     \centering{
%       \footnotesize
%     \begin{tabular}{c | c c c} 
%     \hline
%     Model & PFD  & Identity & PMAE\\
%     \hline\hline
%     StarGANv2 & \textbf{0.51}  & 0.87 & 20.6$^*$\\
%       + one-shot &  \textbf{0.51} & 0.87  & 22.5$^*$\\
%       + pitch &\textbf{0.51}  & 0.86  & \textbf{7.5}\\
%       + PriorGrad (Proposed) &  0.54 & \textbf{0.89} & 8.6 \\
%     \hline
%     \end{tabular}
%     }
% \end{table}
\\
% \subsection{Ablation Study}
\begin{table}[t]
    \caption{\label{tab:subj} {\it Subjective evaluation of one-shot SVC methods on clean data.}}
    \vspace{2mm}
    \centering{
      \footnotesize
    \begin{tabular}{c | c c | c c } 
    \hline
    \multirow{2}{*}{Model} & \multicolumn{2}{c|}{Seen} & \multicolumn{2}{c}{Unseen}\\
        & Natural.  & SIM & Natural.  & SIM\\
    \hline\hline
    Ground Truth &4.65 &4.46 &4.54 &4.38  \\
    \hline
    % AdaINVC+f0 &1.15   &2.29   &1.17   &2.19  \\
    UCDSVC\cite{Polyak20}+enc &2.13 &2.79   &2.17 &2.51 \\
    % \hline
    ROSVC (Ours) &\textbf{3.83}   &\textbf{4.44}   &\textbf{4.08} &\textbf{4.18} \\
    \hline
    \end{tabular}
    }
\end{table}
\\
\textbf{Subjective test}\hspace{2mm}
% \subsection{Subjective test}
We asked 34 audio engineers to rate the naturalness of the audio samples on a scale of 1–5  (1: Bad, 2: Poor, 3:
Fair, 4: Good, 5: Excellent). We also asked them to evaluate the similarity to the reference voice in 1 to 5 scale disregarding the distortion, where 1 indicates completely different and 5 indicates exactly the same singer. We report the mean opinion score (MOS).

The subjective test results on the clean dataset are shown in Table \ref{tab:subj}. We compare the proposed model and UCDSVC, as it is shown to be the most effective among the baselines in the objective evaluation.  As observed, the proposed ROSVC outperforms the baseline by a significant margin in both seen and unseen cases. Moreover, the similarity score of the proposed method on seen singers matches that of the ground truth. Even though the similarity score slightly decreases on unseen singers, it remains high and close to that of the ground truth data. These results confirm that the proposed method is generalized to unseen singers and can generate a singing voice from an unseen singer. 

\begin{table}[t]
    \caption{\label{tab:robust} {\it Subjective metrics on Distorted dataset (Unseen).}}
    \vspace{2mm}
    \centering{
      \footnotesize
    \begin{tabular}{c c c c | c c } 
    \hline
    \multirow{2}{*}{Data} & \multicolumn{2}{c}{Distortion} & \multirow{2}{*}{Model} & \multirow{2}{*}{Natural.}  & \multirow{2}{*}{SIM} \\
        & Src & Ref & & &\\
    \hline\hline
    % Clean & &  & ROSVC (Ours) &4.08   &4.18   \\
    % \hline
    Synthetic &  &   & w/o Robustify &4.08   &4.18   \\
    \hline
    \multirow{2}{*}{Synthetic} & \multirow{2}{*}{\checkmark} &  & w/o Robustify &2.90   &3.29   \\
    & &  & ROSVC (Ours) &\textbf{3.45}   &\textbf{4.00}    \\
    \hline
    \multirow{2}{*}{Synthetic} &  & \multirow{2}{*}{\checkmark} & w/o Robustify &3.86   &2.96   \\
    & &  & ROSVC (Ours) &\textbf{4.45}   &\textbf{3.76}    \\
    \hline
    \multirow{2}{*}{Synthetic} & \multirow{2}{*}{\checkmark} & \multirow{2}{*}{\checkmark} & w/o Robustify &3.24   &3.44   \\
    & &  & ROSVC (Ours) &\textbf{3.85}   &\textbf{4.22}    \\
    \hline
    \multirow{2}{*}{MUSDB18} & \multirow{2}{*}{\checkmark}  & \multirow{2}{*}{\checkmark}  & w/o Robustify &3.54   &3.76    \\
    &   &   & ROSVC (Ours) &\textbf{3.87}   &\textbf{4.22}    \\
    \hline
    \end{tabular}
    }
\end{table}

\begin{table}[t]
   \caption{\label{tab:ablation} {\it Ablation study on seen singers. $^*$ denotes the mPMAE.}}
    \vspace{2mm}
    \centering{
      \footnotesize
    \begin{tabular}{c | c c c} 
    \hline
    Model & PFD  & Identity & PMAE\\
    \hline\hline
    StarGANv2 & \textbf{0.51}  & 0.87 & 20.6$^*$\\
       + one-shot &  \textbf{0.51} & 0.87  & 22.5$^*$\\
       + pitch &\textbf{0.51}  & 0.86  & \textbf{7.5}\\
       + HPG &  0.54 & \textbf{0.89} & 8.6 \\
    \hline
    \end{tabular}
    }
\end{table}

\subsection{Analysis of individual components}
We verify our design choice by progressively adding the proposed components to the baseline StarGANv2 \cite{Li21}. This evaluation is conducted on seen singers because the base StarGANv2 model only supports many-to-many conversion. As shown in Table \ref{tab:ablation}, when we replace the domain-specific style encoder with the domain-independent style encoder, we do not observe much degradation. This suggests that one-shot capability is feasible with no additional cost. However, as suggested by the high mPMAE values, their results sound off-pitch. The proposed pitch conditioning improves the pitch reconstruction accuracy significantly. By replacing the PWG vocoder with HPG, we can observe an improvement in the singer identity similarity and a slight degradation on the PFD and PMAE scores. In our listening test, we found that this degradation is negligible, but the improvement on the singer similarity is significant. Thus, we use HPG for the rest of our experiments.  

\subsection{Robustness against distortions}
\label{sec:ex_dist}
Next, we investigate the robustness against distortions in the singing voice. In the synthetic dataset, we add the reverb and accompaniment music to (i) the reference only, (ii) source only, and (iii) both the source and reference. All singers are unseen during the training in this experiment. We also evaluate the models on non-synthetic data using the MUSDB18 dataset. We manually crop the segments that contain only a single vocal from five songs of different singers and use them for evaluation. 
Note that the evaluation on MUSDB18 dataset is more challenging because MUSDB18 dataset is not used for training and there is a considerable domain gap; vocal tracks in MUSDB18 may contain other types of distortions such as dynamic range compression, and the accompaniment music is musically aligned with the singing voice, meaning that they have the same code progression and vocal tones and are often overlapped with the accompaniments.
% Note that the vocal tracks in MUSDB18 already contain reverb effects and the accompaniment music are is musically aligned with the singing voice, meaning that they have the same code  progression and vocal tones and are often overlapped with the accompaniments, which may be a more challenging setting. 

We conducted a subjective test in the same manner as described in Sec. \ref{sec:ex_osvc} and report the mean opinion scores of naturalness and similarity.
We compare the proposed ROSVC model with the model without \textit{Robustify} (first stage only) to highlight the effect of the second stage training. The results are shown in Table \ref{tab:robust}. The distortions indeed degrade the performance of the model without Robustify. The distortion on the source tends to degrade the naturalness more while the distortion on the reference tends to degrade the similarity. The proposed \textit{Robustify} method significantly improves the subjective scores in all cases, thus demonstrating its effectiveness. It is also worth noting that the naturalness and similarity scores on MUSDB18 match those on Synthetic data. These results indicate that the ROSVC can be generalized to unseen domain data.
% This validates the effectiveness of the propose Robustify approach.

\section{Comparison of neural vocoders}
To further highlight the effectiveness of the proposed hierarchical architecture in a diffusion-based vocoder, we compare the proposed HPG with the baseline diffusion-model, PriorGrad\footnote{\url{https://github.com/microsoft/NeuralSpeech}\label{fn:pg}} \cite{Lee22PriorGrad}. Additionally, we evaluate two GAN-based vocoders: Parallel WaveGAN\footnote{\url{https://github.com/kan-bayashi/ParallelWaveGAN}\label{fn:PWG}} (PWG) \cite{Yamamoto20PWG} and HiFi-GAN\footnote{\url{https://github.com/jik876/hifi-gan}\label{fn:HiFiGAN}} \cite{Kong2020HiFiGAN}. PWG is used as a baseline because recent state-of-the-art text-to-speech models (such as FastSpeech2 \cite{Ren2021FastSpeech2},  DiffSinger \cite{Liu2022DiffSinger}) and voice conversion models (such as StarGANv2-VC \cite{Li21} and ACNN-VC \cite{Um2022ACNNVC}) commonly utilize PWG. All models are trained on the same singing voice dataset by following the instructions in the publicly available implementations$^{\ref{fn:pg}\ref{fn:PWG}\ref{fn:HiFiGAN}}$.

\subsection{Evaluation on voice-converted acoustic features}
We generate 80 voice-converted mel-spectrograms using the proposed ROSVC model by randomly choosing source and reference singing voices. We then generate waveform using different vocoders. 
For subjective evaluation, 20 raters rated the naturalness of samples using headphones on a five-point scale (1: Bad, 2: Poor, 3: Fair, 4: Good, 5: Excellent), and we report the mean opinion score (MOS). 
As for the objective metrics, since there is no ground truth for converted voices, we evaluate voice quality using the predicted MOS (pMOS) introduced in \cite{Mittag2020pMos}. We also evaluated PMAE between the pitch of the target and the converted voice. The results are shown in Table \ref{tab:vc}$^\star$. As we can observe, the proposed model outperforms the baselines in all metrics.
 
% \begin{table}[t]
%     \caption{\label{tab:subj_hpg} {\it MOS results with 95\% confidence interval.}}
%     \vspace{2mm}
%     \centering{
%     % \footnotesize
%     \begin{tabular}{c  c } 
%     \hline
%     \textbf{Model}	&\textbf{MOS}\\
%     \hline
%     Ground Truth    & 4.66 $\pm$ 0.09\\
%     \hline
%     PWG \cite{Yamamoto20PWG}	&2.15 $\pm$ 0.13\\
%     PriorGrad \cite{Lee22PriorGrad}	&3.60 $\pm$ 0.12\\
%     \hline
%     HPG-2 (Ours)	&\textbf{3.95} $\pm$ 0.13\\
%     \hline
%     \end{tabular}
%     }
% \end{table}

\begin{table}[t]
    \caption{\label{tab:vc} {\it Comparison of singing voice vocoders on SVC task.}}
    \vspace{2mm}
    \centering{
    \footnotesize
    \begin{tabular}{c c | c c} 
    \hline
    \textbf{Model}	& \textbf{MOS} [$\uparrow$] &\textbf{pMOS} [$\uparrow$] & \textbf{PMAE} [$\downarrow$]\\
    \hline
    PWG \cite{Yamamoto20PWG}	& 2.81  &3.40   &3.22\\
    HiFi-GAN \cite{Kong2020HiFiGAN} & 2.87  &3.47   &4.39\\
    PriorGrad  \cite{Lee22PriorGrad}	& 3.26  &3.68    &3.25\\
    \hline
    HPG-2 (Ours)	& \textbf{4.07}  &\textbf{3.70}  &\textbf{3.06}\\
    \hline
    \end{tabular}
    }
\end{table}
% We generate two 5-second singing voices for each of the 22 singers for each model and present to 20 raters in random orders.
We compare the MOS of the models in Table \ref{tab:vc}.  Unlike in the speech domain, PWG and HiFi-GAN often suffer from unnatural shakes in pitch when the original singing voice has vibrato, which is possibly one of the reasons for the low MOS. PriorGrad does not produce such unnatural pitch shakes and obtains a higher MOS than PWG. The proposed HPG clearly outperforms all the baselines, providing the best quality. 
We also evaluate a preference score, where raters were asked which of A and B was more natural, with A and B randomly chosen from HPG-2 and PriorGrad. We observe that 85.3\% of the time, raters preferred HPG-2.

\subsection{Evaluation on oracle acoustic features}
We also conducted an objective evaluation with oracle mel-spectrogram features. We use five metrics: (i) real time factor (RTF), measured on a machine with a GeForce RTX 3090 to evaluate the computational cost; (ii) PMAE between the oracle and generated samples, where the pitch is extracted using the WORLD vocoder\cite{WORLD}; (iii) voicing decision error (VDE) \cite{Nakatani2008}, which measures the portion of frames with voicing decision error; (iv) multi-resolution STFT error (MR-STFT) \cite{Yamamoto20PWG}; and (v) Mel cepstral distortion (MCD)\cite{Kubichek1993}. We report the average values of the metrics on the test set.
From the results in Table \ref{tab:obj_hpg}, we can observe that GAN-based methods have higher PMAE values than diffusion-based methods, which is consistent with the subjective observation. In contrast, GAN-based methods obtain lower MR-STFT and MCD values, which is inconsistent with the subjective results. This may be because PWG and HiFi-GAN includes MR-STFT loss or mel-pictogram loss for the training and hence can obtain lower values for the metrics that are related to distortion of the spectrogram. These results suggest that MR-STFT and MCD may be insufficient to evaluate the perceptual quality of a singing voice. 
We also evaluate the effect of increasing the hierarchy to 3-stage and find that HPG-3 further improves the objective metrics with a 30\% increase in computational cost. For a fair comparison, we train a larger PriorGrad model (PriorGrad-L) by increasing the number of layers and channels to $L=40$ and $80$, respectively. PriorGrad-L has the receptive field of 341 ms, which is larger than that of HPG-2 (320 ms) and PriorGrad (256 ms). However, PriorGrad-L performs similarly to PriorGrad and worse than HPGs. These results suggest that the proposed HPG more efficiently scales to the larger model. 

\subsection{The effect of conditioning with lower sampling rate signal}
\label{sec:ex_hpg_condition}
\begin{figure}[t]
  \centering
  \includegraphics[width=\linewidth]{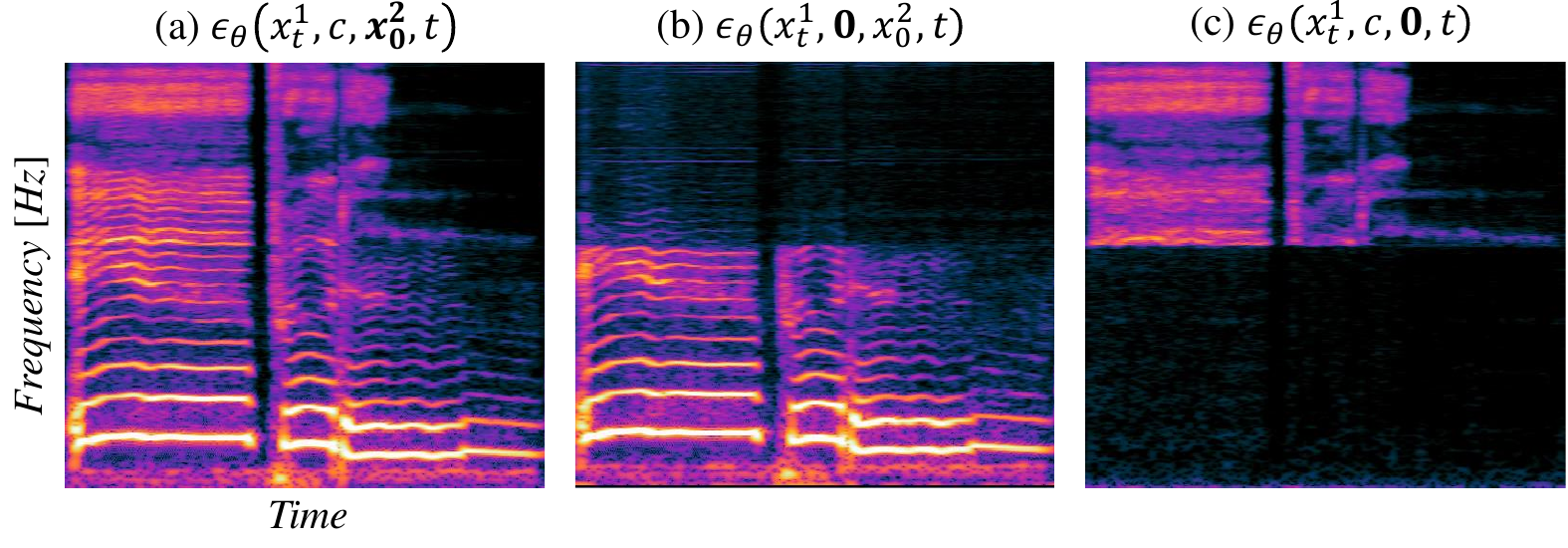}
  \caption{Spectrograms of generated data with different conditioning.}
  \label{fig:condition}
\end{figure}

Finally, we investigate how the HPG model uses the conditioning data. For this, we replace either the mel-spectrogram $c$ or the data at the lower sampling rate $x_0^2$ of the HPG-2 model to zero data during inference and plot the pictograms of the generated the samples. As shown in Figure \ref{fig:condition} (b), the model generates the signal under the Nyquist frequency of the lower module $\frac{f_s^2}{2}$ even when the mel-spectrogram is replaced with zero (b). On the other hand, when the low sampling rate data $x_0^2$ is replaced with zero (Figure \ref{fig:condition} (c)), the model generates only high-frequency components. These results show that the low-frequency components are generated mostly based on $x_0^2$, as expected, even though $c$ contains the information on the low-frequency components, which demonstrates effectiveness of direct access to the waveform at a lower sampling rate. 
Audio samples are available at our website\ref{link:demo}.

\begin{table}[t]
   \caption{\label{tab:obj_hpg} {\it Objective test results. For all metrics, lower is better.}}
    \vspace{2mm}
    \centering{
      \footnotesize
    \begin{tabular}{c | c | c c c c c} 
    \hline
    Model	&RTF &PMAE	&VDE &MR-STFT	&MCD\\
    \hline
    PWG \cite{Yamamoto20PWG}	&0.067	&3.12	&5.61	&1.09	&\textbf{6.63}\\
    HiFi-GAN \cite{Kong2020HiFiGAN}	&\textbf{0.006}	&3.13	&3.91	&\textbf{0.90}	&\textbf{3.84}\\
    PriorGrad \cite{Lee22PriorGrad}	& 0.066	&1.80	&3.96	&1.34	&9.62\\
    PriorGrad-L	&0.093	&2.08	&3.86	&1.38	&9.47\\
    % HiFi-GAN \lbrack \cite{}]	&\textbf{0.006}	&3.13	&3.91	&0.90	&3.84\\
    \hline
    HPG-2 (Ours)	&0.070	&1.82	&3.47	&1.13	&8.97\\
    HPG-3 (Ours)	&0.100	&\textbf{1.67}	&\textbf{3.32}	&1.07	&8.12\\
    \hline
    \end{tabular}
    } 
\end{table}

\section{Conclusion}
We tackled the problem of one-shot SVC with realistic distortion in singing voices, namely, the distortions caused by reverb and accompaniment music. To this end, we proposed a neural network architecture for accurate pitch conditioning, two-stage training for improving the robustness against distortions, and hierarchical diffusion-based neural vocoder to improve the naturalness of singing voices. Objective and subjective experimental results demonstrate that the proposed model outperforms state-of-the-art baselines and is robust to noisy data.

\bibliographystyle{IEEEtran}
\bibliography{MIR,vc,bss,other}

\begin{thebibliography}{10}
\providecommand{\url}[1]{#1}
\csname url@samestyle\endcsname
\providecommand{\newblock}{\relax}
\providecommand{\bibinfo}[2]{#2}
\providecommand{\BIBentrySTDinterwordspacing}{\spaceskip=0pt\relax}
\providecommand{\BIBentryALTinterwordstretchfactor}{4}
\providecommand{\BIBentryALTinterwordspacing}{\spaceskip=\fontdimen2\font plus
\BIBentryALTinterwordstretchfactor\fontdimen3\font minus
  \fontdimen4\font\relax}
\providecommand{\BIBforeignlanguage}[2]{{%
\expandafter\ifx\csname l@#1\endcsname\relax
\typeout{** WARNING: IEEEtran.bst: No hyphenation pattern has been}%
\typeout{** loaded for the language `#1'. Using the pattern for}%
\typeout{** the default language instead.}%
\else
\language=\csname l@#1\endcsname
\fi
#2}}
\providecommand{\BIBdecl}{\relax}
\BIBdecl

\bibitem{Kobayashi14}
K.~Kobayashi, T.~Toda, G.~Neubig, S.~Sakti, and S.~Nakamura, ``Statistical
  singing voice conversion with direct waveform modification based on the
  spectrum differential,'' in \emph{Proc. Interspeech}, 2015.

\bibitem{Nachmani19}
E.~Nachmani and L.~Wolf, ``Unsupervised singing voice conversion,'' in
  \emph{Proc.Interspeech}, 2019.

\bibitem{Deng20}
C.~Deng, C.~Yu, H.~Lu, C.~Weng, and D.~Yu, ``{Pitchnet}: Unsupervised singing
  voice conversion with pitch adversarial network,'' in \emph{Proc. ICASSP},
  2020.

\bibitem{Polyak20}
A.~Polyak, L.~Wolf, Y.~Adi, and Y.~Taigman, ``Unsupervised cross-domain singing
  voice conversion,'' in \emph{Proc. ICASSP}, 2020.

\bibitem{Luo20}
Y.-J. Luo, C.-C. Hsu, K.~Agres, and D.~Herremans, ``Singing voice conversion
  with disentangled representations of singer and vocal technique using
  variational autoencoders,'' in \emph{Proc. ICASSP}, 2020.

\bibitem{Liu21}
S.~Liu, Y.~Cao, D.~Su, and H.~Meng, ``{DiffSVC: A Diffusion Probabilistic Model
  for Singing Voice Conversion},'' in \emph{Proc.IEEE Automatic Speech
  Recognition and Understanding Workshop (ASRU)}, 2021.

\bibitem{Takahashi21}
N.~Takahashi, M.~K. Singh, and Y.~Mitsufuji, ``Hierarchical disentangled
  representation learning for singing voice conversion,'' in \emph{Proc.
  International Joint Conference on Neural Networks (IJCNN)}, 2021.

\bibitem{Liu21FastSVC}
S.~Liu, Y.~Cao, N.~Hu, D.~Su, and H.~Meng, ``{FastSVC: fast cross-domain
  singing voice conversion with feature-wise linear modulation},'' in
  \emph{Proc. ICME}, 2021.

\bibitem{Guo22}
H.~Guo, Z.~Zhou, F.~Meng, and K.~Liu, ``Improving adversarial waveform
  generation based singing voice conversion with harmonic signals,'' in
  \emph{Proc. ICASSP}, 2022.

\bibitem{Zhang22}
Y.~Zhang, P.~Yang, J.~Xiao, Y.~Bai, H.~Che, and X.~Wang, ``K-converter: An
  unsupervised singing voice conversion system,'' in \emph{Proc. ICASSP}, 2022.

\bibitem{Zhou22}
Y.~Zhou and X.~Lu, ``Hifi-svc: Fast high fidelity cross-domain singing voice
  conversion,'' in \emph{Proc. ICASSP}, 2022.

\bibitem{Jayashankar2023}
T.~Jayashankar, J.~Wu, L.~Sari, D.~Kant, V.~Manohar, and Q.~He,
  ``Self-supervised representations for singing voice conversion,'' in
  \emph{Proc. ICASSP}, 2023.

\bibitem{Takahashi2022HPG}
N.~Takahashi, M.~K. Singh, and Y.~Mitsufuji, ``{Hierarchical Diffusion Models
  for Singing Voice Neural Vocoder},'' in \emph{Proc. ICASSP}, 2023.

\bibitem{Villavicencio10}
F.~Villavicencio and J.~Bonada, ``Applying voice conversion to concatenative
  singing-voice synthesis,'' in \emph{Proc. Interspeech}, 2010.

\bibitem{Zhang20}
L.~Zhang, C.~Yu, H.~Lu, C.~Weng, C.~Zhang, Y.~Wu, X.~Xie, Z.~Li, and D.~Yu,
  ``{DurIAN-SC: Duration Informed Attention Network based Singing Voice
  Conversion System},'' in \emph{Proc.Interspeech}, 2020.

\bibitem{Li22}
X.~Li, S.~Liu, and Y.~Shan, ``{A Hierarchical Speaker Representation Framework
  for One-shot Singing Voice Conversion},'' in \emph{Proc. Interspeech}, 2022.

\bibitem{chou2019adainvc}
J.-c. Chou, C.-c. Yeh, and H.-y. Lee, ``One-shot voice conversion by separating
  speaker and content representations with instance normalization,'' in
  \emph{Proc. Interspeech}, 2019.

\bibitem{Wu20VQVCp}
D.-Y. Wu, Y.-H. Chen, and H.-Y. Lee, ``Vqvc+: One-shot voice conversion by
  vector quantization and u-net architecture,'' in \emph{Proc. Interspeech},
  2020.

\bibitem{Chen21AGAINVC}
Y.-H. Chen, D.-Y. Wu, T.-H. Wu, and H.~yi~Lee, ``Again-vc: A one-shot voice
  conversion using activation guidance and adaptive instance normalization,''
  in \emph{Proc. ICASSP}, 2021.

\bibitem{Lin21FragmentVC}
Y.~Y. Lin, C.-M. Chien, J.-H. Lin, H.~yi~Lee, and L.~shan Lee, ``{FragmentVC:
  Any-to-Any Voice Conversion by End-to-End Extracting and Fusing Fine-Grained
  Voice Fragments With Attention},'' in \emph{Proc. ICASSP}, 2021.

\bibitem{Miao20}
X.~Miao, M.~Sun, X.~Zhang, and Y.~Wang, ``{Noise-robust voice conversion using
  high-quefrency boosting via sub-band cepstrum conversion and fusion},''
  \emph{Applied Sciences}, vol.~10, no.~1, 2020.

\bibitem{Mottini21}
A.~Mottini, J.~Lorenzo-Trueba, S.~V.~K. Karlapati, and T.~Drugman, ``{Voicy:
  Zero-Shot Non-Parallel Voice Conversion in Noisy Reverberant Environments},''
  in \emph{Proc. ISCA Speech Synthesis Workshop (SSW)}, 2021.

\bibitem{Xie22}
C.~Xie, Y.-C. Wu, P.~L. Tobing, W.-C. Huang, and T.~Toda, ``{Direct Noisy
  Speech Modeling for Noisy-To-Noisy Voice Conversion},'' in \emph{Proc.
  ICASSP}, 2022.

\bibitem{Aaron2016WN}
A.~van~den Oord, S.~Dieleman, H.~Zen, K.~Simonyan, O.~Vinyals, A.~Graves,
  N.~Kalchbrenner, A.~Senior, and K.~Kavukcuoglu, ``Wavenet: A generative model
  for raw audio,'' \emph{arXiv preprint arXiv:1609.03499}, 2016.

\bibitem{Mehri2017SampleRNN}
S.~Mehri, K.~Kumar, I.~Gulrajani, R.~Kumar, S.~Jain, J.~Sotelo, A.~Courville,
  and Y.~Bengio, ``{SampleRNN: An Unconditional End-to-End Neural Audio
  Generation Model},'' in \emph{Proc. ICLR}, 2017.

\bibitem{Prenger2019WaveGlow}
R.~Prenger, R.~Valle, and B.~Catanzaro, ``{WaveGlow: A flowbased generative
  network for speech synthesis},'' in \emph{Proc. ICASSP}, 2019.

\bibitem{Ping20WaveFlow}
W.~Ping, K.~Peng, K.~Zhao, and Z.~Song, ``{WaveFlow: A compact flow-based model
  for raw audio},'' in \emph{Proc. ICML}, 2020.

\bibitem{Yamamoto20PWG}
R.~Yamamoto, E.~Song, and J.-M. Kim, ``{Parallel WaveGAN: A fast waveform
  generation model based on generative adversarial networks with
  multi-resolution spectrogram},'' in \emph{Proc. ICASSP}, 2020.

\bibitem{Kong21DiffWave}
Z.~Kong, W.~Ping, J.~Huang, K.~Zhao, and B.~Catanzaro, ``{DiffWave: A Versatile
  Diffusion Model for Audio Synthesis},'' in \emph{Proc. ICLR}, 2021.

\bibitem{Kalchbrenner2018WaveRNN}
N.~Kalchbrenner, E.~Elsen, K.~Simonyan, S.~Noury, N.~Casagrande, E.~Lockhart,
  F.~Stimberg, A.~van~den Oord, S.~Dieleman, and K.~Kavukcuoglu, ``Efficient
  neural audio synthesis,'' in \emph{Proc. ICML}, 2018.

\bibitem{Donahue2019WaveGAN}
C.~Donahue, J.~McAuley, and M.~Puckette, ``Adversarial audio synthesis,'' in
  \emph{Proc. ICLR}, 2019.

\bibitem{Kumar2019MelGAN}
K.~Kumar, R.~Kumar, T.~de~Boissiere, L.~Gestin, W.~Z. Teoh, J.~Sotelo,
  A.~de~Brebisson, Y.~Bengio, and A.~Courville, ``{MelGAN: Generative
  Adversarial Networks for Conditional Waveform Synthesi},'' in \emph{Proc.
  NeurIPS}, 2019.

\bibitem{Kong2020HiFiGAN}
J.~Kong, J.~Kim, and J.~Bae, ``Hifi-gan: Generative adversarial networks for
  efficient and high fidelity speech synthesis,'' in \emph{Proc. NeurIPS},
  2020.

\bibitem{Song2019NCSN}
Y.~Song and S.~Ermon, ``Generative modeling by estimating gradients of the data
  distribution,'' in \emph{Proc. NeurIPS}, 2019.

\bibitem{Ho2020DDPM}
P.~A. Jonathan~Ho, Ajay~Jain, ``Denoising diffusion probabilistic models,'' in
  \emph{Proc. NeurIPS}, 2020.

\bibitem{Lu2022cDPM}
Y.-J. Lu, Z.-Q. Wang, S.~Watanabe, A.~Richard, C.~Yu, and Y.~Tsao,
  ``{Conditional Diffusion Probabilistic Model for Speech Enhancement},'' in
  \emph{Proc. ICASSP}, 2022.

\bibitem{Chen2021WaveGrad}
N.~Chen, Y.~Zhang, H.~Zen, R.~J. Weiss, M.~Norouzi, and W.~Chan, ``{WaveGrad:
  Estimating gradients for waveform generation},'' in \emph{Proc. ICLR}, 2021.

\bibitem{Lee22PriorGrad}
S.~gil Lee, H.~Kim, C.~Shin, X.~Tan, C.~Liu, Q.~Meng, T.~Qin, W.~Chen, S.~Yoon,
  and T.-Y. Liu, ``{ "PriorGrad: Improving Conditional Denoising Diffusion
  Models with Data-Dependent Adaptive Prior"},'' in \emph{Proc. ICLR}, 2022.

\bibitem{Koizumi22SpecGrad}
Y.~Koizumi, H.~Zen, K.~Yatabe, N.~Chen, and M.~Bacchiani, ``{SpecGrad:
  Diffusion Probabilistic Model based Neural Vocoder with Adaptive Noise
  Spectral Shaping},'' in \emph{Proc. Interspeech}, 2022.

\bibitem{Li21}
Y.~A. Li, A.~Zare, and N.~Mesgarani, ``{StarGANv2-VC: A Diverse, Unsupervised,
  Non-parallel Framework for Natural-Sounding Voice Conversion},'' in
  \emph{Proc. Interspeech}, 2021.

\bibitem{Choi20StarGANv2}
Y.~Choi, Y.~Uh, J.~Yoo, and J.-W. Ha, ``{StarGAN v2: Diverse Image Synthesis
  for Multiple Domains},'' in \emph{Proc. CVPR}, 2020.

\bibitem{Takahashi21D3Net}
N.~Takahashi and Y.~Mitsufuji, ``Densely connected multidilated convolutional
  networks for dense prediction tasks,'' in \emph{Proc. CVPR}, 2021.

\bibitem{Duan13}
Z.~Duan, H.~Fang, B.~Li, K.~C. Sim, and Y.~Wang, ``{The NUS sung and spoken
  lyrics corpus: A quantitative comparison of singing and speech},'' in
  \emph{Proc. Asia-Pacific Signal and Information Processing Association Annual
  Summit and Conference}, 2013.

\bibitem{Sharma21NHSS}
B.~Sharma, X.~Gao, K.~Vijayan, X.~Tian, and H.~Li, ``{NHSS: A Speech and
  Singing Parallel Database},'' \emph{Speech Communication}, vol.~13, 2021.

\bibitem{sisec2018}
A.~Liutkus, F.-R. St\"{o}ter, and N.~Ito, ``The 2018 signal separation
  evaluation campaign,'' in \emph{Proc LVA/ICA}, 2018.

\bibitem{Yazici2019EMA}
Y.~Yaz{\i}c{\i}, C.-S. Foo, S.~Winkler, K.-H. Yap, G.~Piliouras, and
  V.~Chandrasekhar, ``{The unusual effectiveness of averaging in GAN
  training},'' in \emph{Proc. ICLR}, 2019.

\bibitem{Kingma2015Adam}
D.~P. Kingma and J.~Ba, ``{Adam: A Method for Stochastic Optimization},'' in
  \emph{Proc. ICLR}, 2015.

\bibitem{Kim17HybridASR}
S.~Kim, T.~Hori, and S.~Watanab, ``{Joint CTC-attention based end-to-end speech
  recognition using multi-task learning},'' in \emph{Proc. ICASSP}, 2017.

\bibitem{espnet}
S.~Watanabe, T.~Hori, S.~Karita, T.~Hayashi, J.~Nishitoba, Y.~Unno, N.~E.~Y.
  Soplin, J.~Heymann, M.~Wiesner, N.~Chen, A.~Renduchintala, and T.~Ochiai,
  ``{ESPnet: End-to-End Speech Processing Toolkit},'' in \emph{Proc.
  Interspeech}, 2018, pp. 2207--2211.

\bibitem{Resemblyzer}
``{Resemblyzer},'' \url{https://github.com/resemble-ai/Resemblyzer}.

\bibitem{Wang21}
C.~Wang, Z.~Li, B.~Tang, X.~Yin, Y.~Wan, Y.~Yu, and Z.~Ma, ``{Towards
  High-fidelity Singing Voice Conversion with Acoustic Reference and
  Contrastive Predictive Coding},'' in \emph{Proc. ACM Multimedia}, 2021.

\bibitem{Ren2021FastSpeech2}
Y.~Ren, C.~Hu, X.~Tan, T.~Qin, S.~Zhao, Z.~Zhao, and T.-Y. Liu, ``{FastSpeech
  2: Fast and High-Quality End-to-End Text to Speech},'' in \emph{Proc. ICLR},
  2021.

\bibitem{Liu2022DiffSinger}
J.~Liu, C.~Li, Y.~Ren, F.~Chen, and Z.~Zhao, ``{DiffSinger: Singing Voice
  Synthesis via Shallow Diffusion Mechanism},'' in \emph{Proc. AAAI}, 2022.

\bibitem{Um2022ACNNVC}
J.~S. Um, Y.~Choi, and H.~Kim, ``Acnn-vc: Utilizing adaptive convolution neural
  network for one-shot voice conversion,'' in \emph{Proc. Interspeech}, 2022.

\bibitem{Mittag2020pMos}
G.~Mittag and S.~Moller, ``{Deep Learning Based Assessment of Synthetic Speech
  Naturalness},'' in \emph{Proc. Interspeech}, 2022.

\bibitem{WORLD}
M.~Morise, F.~Yokomori, and K.~Ozawa, ``World: A vocoder-based high-quality
  speech synthesis system for real-time applications,'' \emph{IEICE
  Transactions on Information and Systems}, vol. E99.D, no.~7, pp. 1877--1884,
  2016.

\bibitem{Nakatani2008}
T.~Nakatani, S.~Amano, T.~Irino, K.~Ishizuka, and T.~Kondo, ``A method for
  fundamental frequency estimation and voicing decision: Application to infant
  utterances recorded in real acoustical environments,'' \emph{Speech
  communication}, vol.~50, 2008.

\bibitem{Kubichek1993}
R.~Kubichek, ``{Mel-cepstral distance measure for objective speech quality
  assessment},'' in \emph{Proc. IEEE PACRIM}, 1993.

\end{thebibliography}

\vspace{11pt}
\begin{IEEEbiography}[{\includegraphics[width=1in,height=1.25in,clip,keepaspectratio]{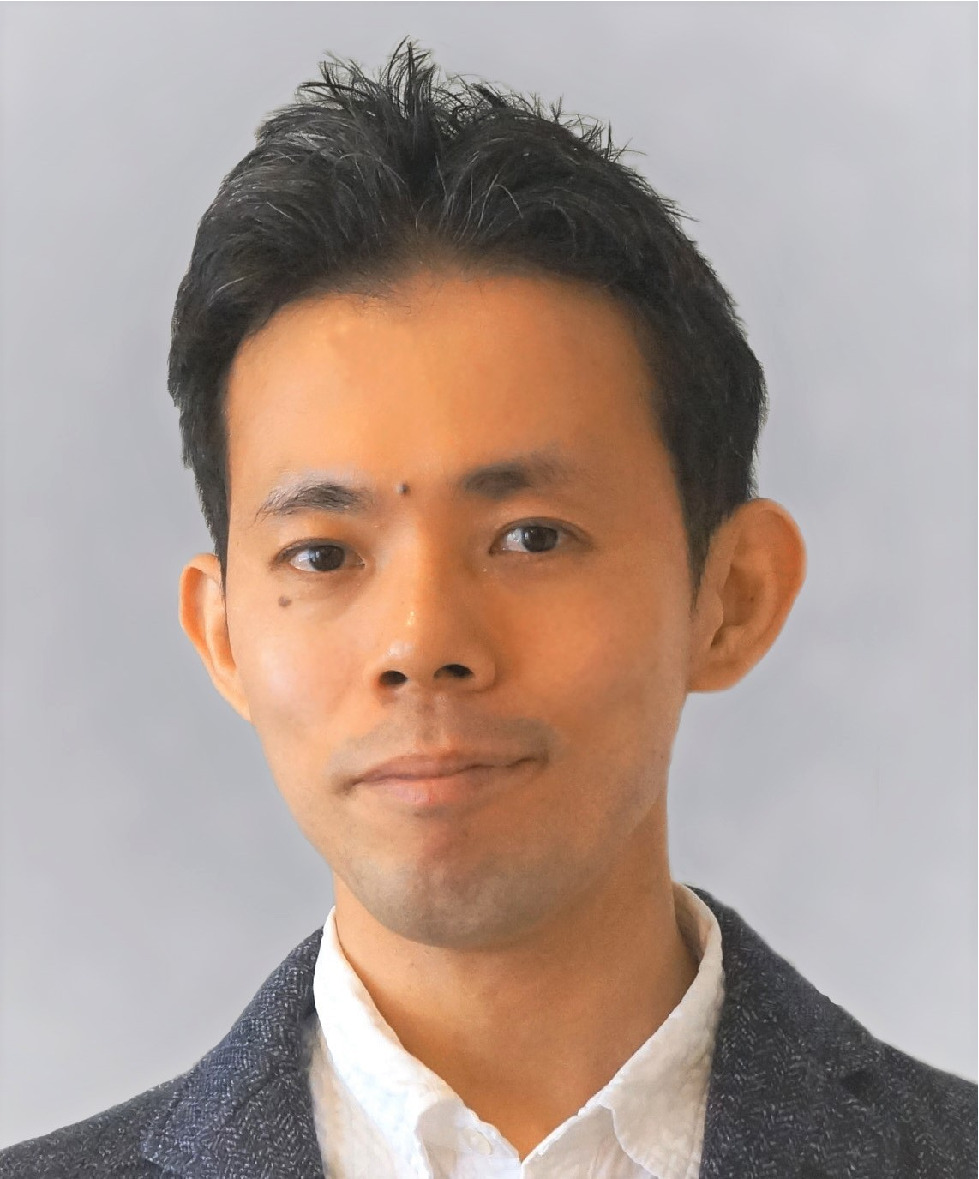}}]{Naoya Takahashi}
received his Ph.D. from University of Tsukuba, Japan, in 2020. Formerly, he had worked at the Computer Vision Lab at ETH Zurich, Switzerland. Since he joined Sony in 2008, he has performed research in the field of audio, computer vision, and machine learning. In 2018, he won the Sony Outstanding Engineer Award, which is the highest form of individual recognition for Sony Group engineers. He has achieved the best scores in several challenges including Signal Separation Evaluation Campaign (SiSEC) 2018, and Detection and Classification of Acoustic Scenes and Events (DCASE) 2021 in Task 3. He has authored several papers and served as a reviewer at ICASSP, Interspeech, CVPR, ICCV, Trans. ASLP, Trans. MM, and more. He co-organized the DCASE 2022 and 2023 task3.
\end{IEEEbiography}

\begin{IEEEbiography}[{\includegraphics[width=1in,height=1.25in,clip,keepaspectratio]{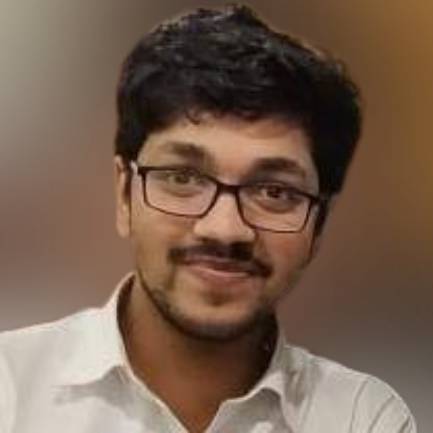}}]{Mayank Kumar Singh}
received his BTech from the Indian Institute of Technology, Bombay, India in 2020, where he worked in the Vision and Image Processing (VIP) Lab and Medical Deep Learning and Artificial Intelligence Lab (MeDAL). Since he joined Sony in 2020, he has performed research in the field of audio, computer vision and machine learning. He has authored several papers at ICASSP, Interspeech and IJCNN.
\end{IEEEbiography}

\begin{IEEEbiography}[{\includegraphics[width=1in,height=1.25in,clip,keepaspectratio]{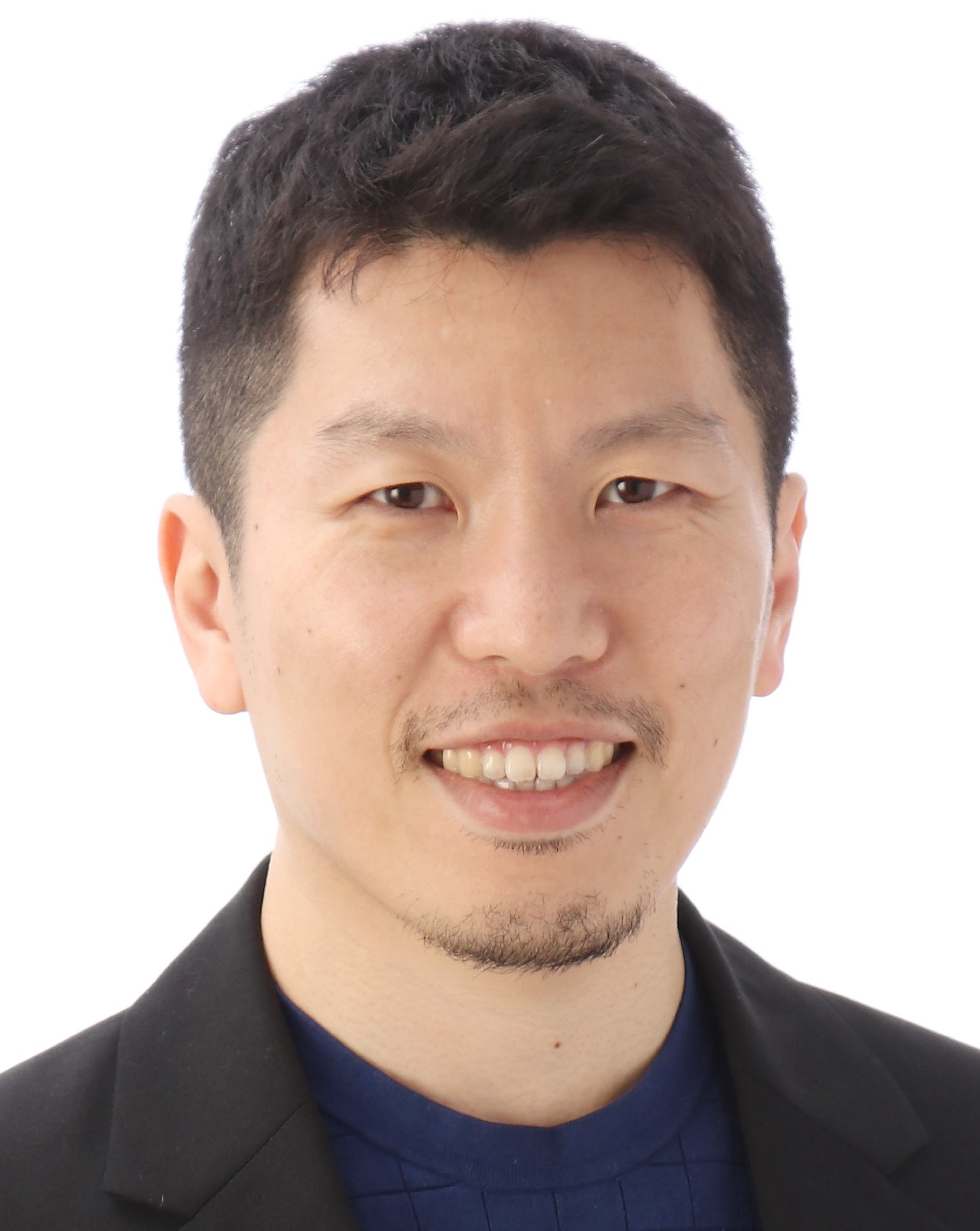}}]{Yuki Mitsufuji} received B.S. and M.S. degrees in information science from Keio University in 2002 and 2004, respectively, and obtained a Ph.D. in information science and technology from the University of Tokyo in 2020. Currently, he is leading the Creative AI Lab at Sony Group Corporation while serving as a Specially Appointed Associate Professor at the Tokyo Institute of Technology. Since joining Sony Corporation in 2004 , he has led teams that developed the sound design for the PlayStation Gran Turismo Sport game and a spatial audio solution called Sonic Surf VR. He has also won several awards, including a TIGA award for best audio design (for Gran Turismo Sports) and a jury selection at the Japan Media Arts Festival for their 576-channel sound field synthesis called Acoustic Vessel Odyssey. From 2011 to 2012, he was a visiting researcher at the Analysis/Synthesis Team, Institut de Rechereche et Coordination Acoustique/Musique (IRCAM), Paris, France. He was involved in the 3DTV content search project sponsored by European Project FP7, in research collaboration with IRCAM. In 2021, his team organized the Music Demixing (MDX) Challenge, where Sony Music provided a professionally produced music dataset for the evaluation of submitted systems to an online platform on AIcrowd. His team also participated in the DCASE2021 Challenge and achieved first place in Task 3.
\end{IEEEbiography}

\vspace{11pt}


\vfill

\end{document}